\documentclass[12pt]{article}
\usepackage{amsfonts}
\usepackage{graphicx}\usepackage{color}\usepackage[utf8]{inputenc}
\usepackage{amsmath,amssymb}
\makeatletter \@addtoreset{equation}{section}\makeatother\def\Z{\mathbb Z}
\def\Z{\mathbb Z}
\def\cW{\mathcal{W}}

\def\A{\mathbb A}
\def\W{\mathbb W}

\def\P{\mathbb P}

\def\be{\begin{equation}}\def\ee{\end{equation}}\def\f{\frac}
\def\A{\mathbb A}\def\bea{\begin{eqnarray}}\def\eea{\end{eqnarray}}
\def\ben{\begin{displaymath}}
\def\ba{\begin{array}{c}}\def\bal{\begin{array}{l}}\def\ea{\end{array}}\def\sh{\mathrm{sh}}\def\ch{\mathrm{ch}}
\def\Z{\mathbb Z}
\def\cC{{\mathcal{C}}}\def\cH{{\mathcal{H}}}

\def\een{\end{displaymath}}

\begin{document}

\title{
Transmutations of supersymmetry  through 
soliton scattering, and self-consistent condensates
}

\author{Adri\'an Arancibia and Mikhail S. Plyushchay  \\
[4pt]
{\small \textit{
Departamento de F\'{\i}sica,
Universidad de Santiago de Chile, Casilla 307, Santiago 2,
Chile  }}\\
 \sl{\small{E-mails: 
adaran.phi@gmail.com, mikhail.plyushchay@usach.cl
}}
}
\date{}
\maketitle

\begin{abstract}
We consider the two most general 
families of the   $(1+1)$D  Dirac systems with 
transparent scalar  potentials,  
and two related  families of the paired  
reflectionless Schr\"odinger operators. 
The ordinary $\mathcal{N}=2$ supersymmetry for 
such Schr\"odinger  pairs is 
 enlarged up to an exotic  $\mathcal{N}=4$ nonlinear
centrally extended supersymmetric structure, which
involves  two bosonic 
integrals composed from the Lax-Novikov operators for the 
stationary Korteweg-de Vries  hierarchy.   Each associated single
 Dirac system 
displays  a proper $\mathcal{N}=2$  nonlinear supersymmetry
with a non-standard grading operator.
One of the two families of the  first and second order systems 
exhibits the unbroken  supersymmetry, while another is described 
by  the broken exotic supersymmetry. 
The two families are shown to be mutually transmuted  
by applying a certain limit procedure
to the soliton scattering data.
We relate the  topologically trivial and 
nontrivial transparent potentials
with self-consistent inhomogeneous condensates 
in Bogoliubov-de Gennes
and Gross-Neveu models, and indicate the exotic 
$\mathcal{N}=4$ nonlinear supersymmetry 
of the paired  reflectionless Dirac systems.
\end{abstract}

\section{Introduction}

The 
Schr\"odinger and  Dirac equations 
with reflectionless, or soliton  potentials 
are exactly solvable.  The reflectionless 
potentials of a general form  for one-dimensional 
Schr\"odinger equation
 were obtained for
 the first time  by Kay and Moses 
 by solving the problem 
of a theoretical construction of a solid dielectric medium 
which is perfectly transparent to electromagnetic radiation  \cite{KayMos}. 
Such perfectly transparent potentials appear in 
(1+1)-dimensional  
Gross-Neveu (GN) model  \cite{GrNe}--\cite{Feinberg}, 
and are closely related with 
a nonlinear problem of self-consistency 
of the  Bogoliubov-de Gennes (BdG)  equations 
\cite{NNB,deG,Stone}.
They find applications in the description of a 
broad spectrum of phenomena in diverse areas of physics 
such as  conducting polymers
  \cite{TLM}--\cite{heeger},  charge fractionalization 
   \cite{Jackiw}--\cite{Jack84},  and superconductivity 
   \cite{NNB}--\cite{Stone},
\cite{ThiesM}--\cite{Andr},  just to mention a few.
There is also a great interest in supersymmetry 
associated with fermions in 
soliton backgrounds 
 \cite{WitOl}--\cite{Shifman:1998zy}.

Reflectionless potentials play 
a fundamental role in 
the theory of integrable systems. 
They appear as soliton solutions, particularly, 
 to the 
Korteweg-de Vries  (KdV) and modified Korteweg-de Vries (mKdV)
 equations.  
 Their explicit form can be obtained
 by means of inverse scattering method, 
 by B\"acklund transformation, or by Darboux-Crum transformations
 \cite{KayMos}, \cite{MatSal}--\cite{BurChau}.
A characteristic feature of the two last methods is a possibility to 
construct these potentials 
from simple (formal) solutions of the free particle.

In the present  work we focus on 
 the Darboux  transformations. 
 In this picture there appear 
 the first order differential  operators,
  which
  intertwine  reflectionless 
  Schr\"odinger,    and perfectly transparent 
  Dirac Hamiltonians.  This  
  will allow us, following the line of 
  refs. \cite{PAM,PAM+,PAN,PlyNi,MPhid,CJP,FDP}, 
  to study the interrelations between 
  the exotic nonlinear supersymmetric structures emerging 
  in  the first and in the second order quantum  reflectionless
 systems of  the most general form corresponding 
 to the KdV and mKdV solitons
\footnote{For the earlier studies
related to the appearance  of the exotic extended  supersymmetric 
structure in such class of the systems characterized by the 
presence of the nontrivial Lax-Novikov integral,
see also Refs. 
\cite{Ges1,VesSh,GesWei,FMROS1,ASo}.  }.
  We also will observe an interesting phenomenon 
 of  transmutation of supersymmetry 
 associated with  the  soliton scattering,
 and  will  relate the construction  to the self-consistent inhomogeneous 
 condensates appearing in the GN and BdG models.

A  relation of the soliton potentials  with the GN model 
 \cite{GrNe} goes back to  the
 famous result of 
Dashen, Hasslacher and Neveu \cite{DaHN}, who found that minimizing 
the effective action of the 
model for the `condensate function'  $\sigma(x)=-g\bar{\psi}\psi$, 
results in the condition that the Schr\"odinger potentials 
$U_{\pm}(x)$ 
given in terms of the Miura transformation \cite{Miura},
$U_{\pm}(x)\equiv g^2\sigma^2(x)\pm g{d\sigma(x)}/{dx}$,  
  have to be reflectionless. 
On the other hand, the Dirac system with transparent potential $\sigma(x)$
appears in  the Takayama--Lin-Liu--Maki  (TLM) model for conducting polymers   
   \cite{TLM}, which 
 is a continuous model for solitons in  
 polyacetylene, where the  kink and  kink-antikink solutions were found
\cite{kimlee}.
Though these two models have distinct physical interpretations, 
they are equivalent mathematically,
and the methods developed in the context of the GN model were  
applied in the study of the 
 TLM model \cite{CamBi,BrKir,heeger}.
In general, the self-consistent  solutions of the GN model are related with 
the Ablowitz-Kaup-Newell-Segur  hierarchy  \cite{FDP,TakTYN}, 
and by the same reason are intimately related with 
integrable systems in  1+1 dimensions. Particularly, some solutions to the 
GN model were found to be related with the breather type solutions of the sinh-Gordon and 
 nonlinear Schr\"odinger   equations \cite{KlTh}.

The integrability of the equations of the KdV and mKdV hierarchies 
can be associated with existence of an 
auxiliary spectral problem
given in terms of the spectral operator 
$H$ and the evolution generator $P_j$. 
The consistency condition appears there 
in the form of the equation for the Lax pair $(H,\, P_j)$,
 $\frac{d H}{d t_j}=[H,P_j]$, which
 is equivalent to the 
corresponding  evolution 
equation. 
For the  $j$-th equation of the KdV hierarchy, 
 $H=-\frac{d^2}{dx^2}+U$ 
 is the  Schr\"odinger operator,
 while  $P_j$, $j=0,1,\ldots,$ 
  is an anti-Hermitian monic differential operator 
  of the 
  form $P_j=\frac{d^{2j+1}}{dx^{2j+1}}
  +a_{2j-1}\frac{d^{2j-1}}{dx^{2j-1}}+\ldots 
  +a_0$ with coefficient functions $a_i$ given 
  in terms of the potential  $U$ and its $x$-derivatives. 
 The case of the KdV equation corresponds to $j=1$,
 and its an $n$-soliton solution $U_n(x,t)$ satisfies simultaneously 
 the equation $[\mathcal{L}_n,H_n]=0$, which is the 
  nonlinear ordinary
 differential equation of order $2n+1$ in $x$ variable.
This is the $n$-th stationary  equation for the KdV
hierarchy, in which  $t_1=t$ plays a role of an external 
 parameter. 
The operator 
$\mathcal{L}_n={P}_{n}+
\sum_{j=0}^{n-1}c_j {P}_j$, where $c_j$ are some real coefficients,
 is  the Lax-Novikov
 nontrivial integral of motion for  $H_n=-\frac{d^2}{dx^2}+U_n$
  \cite{SPN,NovZak,GesWei} .  
  According to a celebrated result of 
  Burchnall and Chaundy  \cite{BurChau},  the square  
  of the order $2n+1$ differential operator 
$\mathcal{L}_n$ reduces to a certain  
polynomial  in $H_n$.
\vskip0.1cm

One can  construct the pair 
 ($H_n$,  $\mathcal{L}_n$)  corresponding to 
 an  $n$-soliton potential $U_n$ 
 recursively, staring from the free particle case with 
 $H_0=-\frac{d^2}{dx^2}$
  and $\mathcal{L}_0=\frac{d}{dx}$ ($U_0=0$),
 and using the Darboux transformations. 
 If we restrict ourselves by  regular on the $x$-axis  potentials, 
 then  at  each step,  
\begin{itemize}

\item {\bf i)}  from    $U_n$,  we construct an almost isospectral
reflectionless  potential  $U_{n+1}$ 
with one more bound state in comparison with 
$U_{n}$, and from  the  Lax-Novikov integral $\mathcal{L}_{n}$  for $H_n=-\frac{d^2}{dx^2}
+U_n$, we  obtain the integral 
$\mathcal{L}_{n+1}$ for reflectionless Schr\"odinger system 
$H_{n+1}$.

\end{itemize}
The interesting point here is that 
having reflectionless Schr\"odinger potential $U_n$ of a general form,
by applying Darboux transformation of another nature,

\begin{itemize}

\item {\bf ii)}  we can construct from  $U_n$
another $n$-soliton potential  $\tilde{U}_n$ to be 
completely isospectral  to $U_n$, 
and from  $\mathcal{L}_n$ we can obtain the corresponding integral  
$\tilde{\mathcal{L}}_n$ for $\tilde{H}_n$.

\end{itemize}
The latter construction can be realized by applying a certain 
limit procedure for soliton scattering data of the reflectionless potential 
$U_{n+1}$.  By a similar limit procedure one can also relate 
$\tilde{U}_n$ with $U_{n-1}$,  and  $U_n$ with $U_{n-2}$.  
In both cases i) and ii) above, one can associate with 
the corresponding pairs 
of the  reflectionless second order Hamiltonians the exotic 
$\mathcal{N}=4$ nonlinear supersymmetry 
that includes two bosonic integrals composed from
Lax-Novikov integrals for the partner subsystems.

Exploiting the knowledge of the Darboux transformations for the KdV, 
one can generalize the construction for the case of the 
mKdV to get the  transparent 
Dirac systems with the multi-kink scalar potentials, and to identify 
for  each such a single 
first order matrix system a proper exotic $\mathcal{N}=2$ supersymmetry. 
As in the Schr\"odinger case, the transparent Dirac multi-kink potentials
are separated into the two groups:
one of them  is formed by topological and another by 
non-topological  scalar potentials. 
The topological potentials are associated with the case i) above,
 and represent  the configurations of $n$ kinks 
and $n\pm 1$ antikinks. The  non-topological transparent potentials  
correspond to the case ii),  and 
represent  a certain superposition of $n$ kinks and $n$ antikinks. 
We shall show  how the kinks and antikinks in transparent Dirac potentials 
gather together in such a way that 
their amplitudes and phases are fixed by supersymmetry of the 
paired  reflectionless Schr\"odinger  systems.

\vskip0.1cm

The paper  is organized as follows. 
In Section 2 we review shortly  the recursive 
construction of the multi-soliton Schr\"odinger potentials 
of the most general form 
in terms of the Darboux transformations,
and describe the spectra 
of the corresponding reflectionless 
Schr\"odinger operators. 
We identify there 
the independent  differential  operators of orders $1$ and $2n$,
which intertwine the neighbour in a recursive chain 
Schr\"odinger Hamiltonians 
$H_n$ and $H_{n-1}$, and find 
the Lax-Novikov integral of differential order $2n+1$ for $H_n$.
In Section 3 we describe a unique another family of the reflectionless
pairs ($H_n$, $\tilde{H}_n$) with completely 
isospectral partners,  which are also intertwined  by the 
Darboux generators to be differential operators of the same orders 
$1$ and $2n$, and find
a certain limit procedure, related to the soliton scattering, which 
mutually transmutes the two indicated families of the 
pairs of the transparent Schr\"odinger systems. 
In Section 4 we study  the exotic nonlinear supersymmetries 
of the two families of the Schr\"odinger systems composed from the 
reflectionless isospectral, ($H_n$, $\tilde{H}_n$), 
and almost isospectral,  ($H_n$,  $H_{n-1}$),
pairs, and observe the  transmutations between these two families 
through the soliton scattering.  
In Section 5 we study  the transparent Dirac systems 
associated with the two families of the superextended 
reflectionless Schr\"odinger systems, where we show 
that each single transparent Dirac system possesses 
its own exotic nonlinear supersymmetry. 
The last Section 6 is devoted to the discussion of the obtained results
and outlook. There we relate  the perfectly transparent scalar 
Dirac potentials with the 
self-consistent inhomogeneous condensates 
appearing in the BdG and GN models, and indicate 
the exotic  $\mathcal{N}=4$ nonlinear supersymmetry 
of the paired  reflectionless Dirac systems.

\vskip0.3cm

\section{Reflectionless Schr\"odinger potentials 
and  Darboux-Crum transformations}

Let 
$H_n=-d^2/dx^2 +U_n(x)$ 
be a reflectionless Schr\"odinger system 
with a potential of  a general $2n$-parametric form 
 $U_n(x)=U_n(x;\kappa_1,\tau_1,
 \ldots, \kappa_n,\tau_n)$ such that 
 $U_n(x)\rightarrow 0$ for $x\rightarrow\pm\infty$.
 Parameters $\kappa_j$, $j=1,2,\ldots,n$,
 $0<\kappa_1<\ldots<\kappa_n$, 
 correspond to  the  energy levels   of the $n$ bound states,
 $E_j=-\kappa_j^2$. They also define the transmission amplitude
in the scattering sector with $E=k^2\geq 0$\,: 
$t(k)=\prod_{j=1}^{n}\frac{k+i\kappa_j}{k-i\kappa_j}$, and so
$\vert t(k) \vert=1$ for any real value of the wave number $k$.
The parameters $\tau_j$ are related to the norming constants 
of the bound states solutions \cite{NovZak,El}, and 
their variation  provides  an isospectral deformation 
of the quantum system. 

{}From the viewpoint of the inverse scattering method,
function $U_n(x;\kappa_1,\tau_1,
 \ldots, \kappa_n,\tau_n)$ corresponds to the instant image of
  the $n$-soliton solution $U_n(x,t)$  to the
  KdV equation  $u_t-6uu_x+u_{xxx}=0$.
  For large negative and positive values of time $t$
  the  $U_n(x,t)$ can be represented as a superposition 
  of  $n$ one-soliton solutions of the amplitudes
  $2\kappa_j^2$ propagating to the right 
  at the speeds $v_j=4\kappa_j^2$,
  \begin{equation}\label{Unxt}
 	 U_n(x,t)=-\sum_{j=1}^n 
  	2\kappa_j^2\, \text{sech}^2 \kappa_j(x-4\kappa_j^2t-\varphi^\pm_j)\quad
 	 \text{for}\quad  t\rightarrow \pm\infty\,.
  \end{equation}
 The  phases  $\varphi^\pm_j$ defined for $t\rightarrow \pm \infty$
 are given by \cite{NovZak,El}
  \begin{equation}\label{Deltaphin}
	\varphi_l^\pm=\tau_l^0 \pm \frac{1}{2\kappa_l}
	\left\{
	\sum_{j=l+1}^n\log\bigg| \frac{\kappa_l+\kappa_j}{
	\kappa_l-\kappa_j}\bigg| -
	\sum_{j=1}^{l-1}\log\bigg|\frac{\kappa_l+\kappa_j}{
	\kappa_l-\kappa_j}\bigg|\right\}\,,
\end{equation}
where it is implied that  for $l=n$  and  $l=1$  the first and, respectively, 
the second sum disappears. The parameter  
$\tau_l^0$ corresponds  to the mean
of the asymptotic phases,
$\tau_l^0=\frac{1}{2}(\varphi_l^+ + \varphi_l^-)$.
According to (\ref{Deltaphin}),  the  solitons demonstrate 
in some sense a fermion-like behaviour\,:
$\vert\varphi_l^\pm\vert,\, \vert\varphi_{l+1}^\pm
\vert\rightarrow \infty$ as soon as $\kappa_l\rightarrow\kappa_{l+1}$.
In the two-soliton case, (\ref{Deltaphin}) gives\,:
\begin{equation}\label{phasen=2}
	\varphi_1^\pm=\tau_1^0 \pm \frac{1}{2\kappa_1}
	\log\bigg| \frac{\kappa_1+\kappa_2}{
	\kappa_1-\kappa_2}\bigg|\,,\qquad
	\varphi_2^\pm=\tau_2^0 \mp \frac{1}{2\kappa_2}
	\log\bigg| \frac{\kappa_1+\kappa_2}{
	\kappa_1-\kappa_2}\bigg|\,.
\end{equation}
\vskip0.1cm

Our consideration  will be based 
on  the method of iterated Darboux transformations (or,
that is the same, the Darboux-Crum transformations)
\cite{MatSal}, by which  
the quantum mechanical reflectionless system with $n$ 
bound states can be constructed  from a free 
particle  system  $H_0=-\frac{d^2}{dx^2}$\,,
\begin{equation}\label{lnW}
    H_n=H_0+U_n(x),\qquad
    U_n=-2\frac{d^2}{dx^2}\log \W_n\,.
\end{equation}
Here  $\W_n$ is the Wronskian of  $n$ 
formal (non-physical) eigenstates  $\psi_j$ of  $H_0$,
$H_0\psi_j=-\kappa_j^2\psi_j$, 
\begin{equation}\label{Wron-n}
   \W_n= \W(\psi_1,\ldots,\psi_n)=\det \mathcal{A},\qquad
    \mathcal{A}_{ij}=\frac{d^{i-1}}{dx^{i-1}}\psi_j,\qquad
    i,j=1,\ldots, n\,,
\end{equation}
which are chosen as follows,
\begin{equation}\label{psij}
    \psi_j=\left\{\begin{matrix}\cosh (\kappa_j(x+\tau_j))&
   \text{for}\,\,\,  j={\rm odd}\\ \sinh
    (\kappa_j(x+\tau_j))& \text{for}\,\,\,
     j={\rm even}\end{matrix}\,\,,\right.
    \qquad 0<\kappa_1<\kappa_2<...<\kappa_{j-1}<\kappa_n\,.
\end{equation}
Eigenfunctions  $\Psi_0(x;E)\neq \psi_j$  of $H_0$,
$H_0 \Psi_0(x;E)=E\Psi_0(x;E)$, are mapped into
 the eigenfunctions 
$\Psi_n(x;E)$ of  $H_n$, $H_n \Psi_n(x;E)=E\Psi_n(x;E)$,
by means of  the Wronskian fractions,
\begin{equation}\label{PsinE}
\Psi_n(x;E)=\W(\psi_1,\ldots,\psi_{n},\Psi_0(E))/\W_{n}\,.
\end{equation}
The eigenfunctions in 
the scattering sector with $E=k^2\geq 0$,
$k\geq 0$, 
and (not normalized) bound states with energies $E_j=-\kappa_j^2$,
$j=1,\ldots,n$, of the system $H_n$ 
are given  then by the relations
\begin{equation}\label{est}
	\Psi^\pm_n(k^2)=\W(\psi_1,\ldots,\psi_{n},e^{\pm ikx})/\W_{n}\,
	,\qquad
	\Psi_n(-\kappa_j^2)=\W(\psi_1,\ldots,\psi_{n},\f{d\psi_j}{dx})/\W_{n}\,.
\end{equation}
The derivative
$\f{d\psi_j}{dx}$ is a non-physical eigenfunction of $H_0$ 
which is   linearly  independent  from  the corresponding 
non-physical eigenfunction $\psi_j$ from (\ref{psij}).

Coherently with (\ref{lnW}), we put $\W_0=1$ and 
define the \emph{prepotentials} $\Omega_n$, $n=0,1,\ldots$, 
\begin{equation}\label{PrepOm}
	\Omega_n=-\frac{d}{dx}\log \W_n\qquad
	\Rightarrow\qquad
	 \f{d}{dx}\Omega_n
    =\frac{1}{2}U_n\,.
\end{equation}
Then we introduce the first order differential operators 
 \begin{equation}\label{An}
    A_n=
    \f{d}{dx}+\mathcal{W}_n\,,\qquad
    \mathcal{W}_n=
    \Omega_n-\Omega_{n-1}\,,
\end{equation}
where, particularly, 
$\mathcal{W}_1=\Omega_1=-
	\kappa_1\tanh\kappa_1(x+\tau_1)\,.
$
These  operators and their conjugate ones 
factorize  the reflectionless systems  
 $H_{n-1}$  and  $H_n$  
having the  $(n-1)$- and $n$-soliton potentials
$U_{n-1}(x;\kappa_1,\tau_1,
 \ldots, \kappa_{n-1},\tau_{n-1})$
 and 
$U_n(x;\kappa_1,\tau_1,
 \ldots, \kappa_{n-1},\tau_{n-1},\kappa_n,\tau_n)$, 
\begin{equation}\label{fact}
A_n^\dag A_n=
    H_{n-1}+\kappa^2_{n},\qquad
    A_nA_n^\dag=
    H_n+\kappa^2_{n},   
\end{equation}
and  intertwine them, 
\begin{equation}\label{inter1}
	 A_nH_{n-1}=H_nA_n\,,\qquad
	 A_n^\dagger H_n=H_{n-1}A_n^\dagger\,.
\end{equation}
The operator  $A_n$  can be presented
equivalently as $A_n=\Psi_{n-1}^A\frac{d}{dx}(1/\Psi_{n-1}^A)$,
where $\Psi_{n-1}^A\equiv\f{\W_{n}}{\W_{n-1}}$
is a nodeless non-physical eigenfunction
 of $H_{n-1}$,
$H_{n-1}\Psi_{n-1}^A=-\kappa_{n}^2\Psi_{n-1}^A$. 
This function  is  a formal (exponentially blowing up
at $x=\pm\infty$)  
zero mode of  the first order differential operator $A_n$,
$A_n\Psi_{n-1}^A=0$.
Any other (physical or non-physical) eigenstate 
$\Psi_{n-1}(E)$  of $H_{n-1}$, 
$H_{n-1}\Psi_{n-1}(E)=E\Psi_{n-1}(E)$,
is mapped by $A_n$  into the eigenstate 
of $H_n$, 
\begin{equation}\label{AnPsi}
  \Psi_n(E)=A_n\Psi_{n-1}(E)\,,
\end{equation}
with the same eigenvalue, 
$H_{n}\Psi_{n}(E)=E\Psi_{n}(E)$.

By iteration of (\ref{inter1}), reflectionless system 
$H_n$ can be related with the free particle $H_0$, 
\begin{equation}\label{HnAn}
    \A_nH_0=
    H_n\A_n\,,\qquad
    \A_n^\dagger
    H_n=H_0\A_n^\dagger\,,
\end{equation}
where $\A_n$ is the differential 
operador of order $n$,
\begin{equation}\label{AnCD}
    \A_n\equiv A_n\ldots
    A_1\,.
\end{equation}
In terms of (\ref{AnCD}), we  define  the differential operator 
of order $2n$,
\begin{eqnarray}\label{B2n}
	&\mathcal{B}_1=
	A_1\left( -\f{d}{dx}+\kappa_1\right)\,, \qquad
	\mathcal{B}_n=
	\A_n \left(-\f{d}{dx}+\kappa_n\right)\A^\dag_{n-1}\quad
	\text{for}\quad n=2,\ldots\,.&
\end{eqnarray}
The iteration of relations (\ref{inter1})   shows that $\mathcal{B}_n$ 
and $\mathcal{B}_n^\dagger$
also intertwine reflectionless Hamiltonians $H_n$  and $H_{n-1}$,
\begin{equation}\label{Bninter}
	 \mathcal{B}_nH_{n-1}=H_n\mathcal{B}_n\,,\qquad
	  \mathcal{B}_n^\dagger H_n=H_{n-1} \mathcal{B}_n^\dagger\,.
\end{equation}
Unlike $A_n$ and  $A_n^\dagger$, they do this not directly
but via the `virtual' free particle system $H_0$, for which 
the first order differential operator $\f{d}{dx}$,
appearing explicitly in the structure of $\mathcal{B}_n$  is an  
integral of motion.
Instead of
(\ref{fact}) we have the relations
\begin{equation}\label{BBHn}
	\mathcal{B}_n\mathcal{B}_n^\dagger=
	\prod_{j=1}^{n} (H_n+\kappa_j^2)^2\,,\qquad
	\mathcal{B}_n^\dagger \mathcal{B}_n
	=\prod_{j=1}^{n} (H_{n-1}+\kappa_j^2)^2\,.
\end{equation}

The operator (\ref{AnCD}) also allows us to find a nontrivial 
integral 
for reflectionless system $H_n$,
\begin{equation}\label{Pndef}
    \mathcal{L}_{n}=
    \A_n p
    \A_n^\dagger\,,\qquad
    \mathcal{L}_{n}^\dagger= \mathcal{L}_{n}\,,\qquad
    [ \mathcal{L}_{n},H_n]=0\,.
\end{equation}
This  differential operator of order $2n+1$ is the Lax-Novikov integral
for  the  $H_n$. It is a 
Darboux-dressed form of the integral $p=-i\frac{d}{dx}$ for the free
particle  system
$H_0$, which satisfies the nonlinear supersymmetry type relation
\begin{equation}\label{LHspectral}
	 \mathcal{L}_{n}^2=H_n\prod_{i=1}^n(H_n+\kappa_i^2)^2\,.
\end{equation}
The property of commutativity of $\mathcal{L}_n$ with $H_n$ 
means that the potential 
 $U_n=2\frac{d}{dx}\Omega_n$ 
 is a solution of the $n$-th member of the KdV
 stationary  hierarchy~\footnote{Note that unlike Section 1, we take 
 $\mathcal{L}_n$ here  in a Hermitian form.}.

Using analogs of the integrals (\ref{Pndef}) for  $H_l$ with $0<l<n$,
one could try to  construct the operators intertwining $H_{n-1}$ and $H_{n}$
with $n>1$  via a virtual $H_l$ system.
In such a way we obtain, however,  a combination of
$\mathcal{B}_n$ and $A_n$ with a coefficient before 
the latter operator to be a polynomial of order $(n-1)$ in $H_{n-1}$.
For instance,  $-iA_n\mathcal{L}_{n-1}$ is the differential operator 
of order $2n$, which, like $\mathcal{B}_n$, intertwines 
$H_{n-1}$ with $H_n$, but reduces to  $-iA_n\mathcal{L}_{n-1}=\mathcal{B}_n
-\kappa_n A_n\prod_{i=1}^{n-1}(H_{n-1}+\kappa_i^2)$,
and so, is not a new, independent intertwining operator.
At the same time  note that the 
intertwining operators $A_n$ and $\mathcal{B}_n$,
and the integral $\mathcal{L}_n$ are related  with the Hamiltonian $H_n$
 by
the identity
\begin{equation}\label{ABLHid}
 \mathcal{B}_nA_n^\dagger +i\mathcal{L}_n=
\kappa_n\prod_{i=1}^n(H_n+\kappa_i^2).
\end{equation}

In conclusion of this section 
 it is worth stressing once more that the existence of the nontrivial, 
 order  $2n$  intertwining 
 operator  $\mathcal{B}_n$ in addition to the first order
 Darboux generator $A_n$ as well as of the order $2n+1$ 
 integral 
 $\mathcal{L}_n$  originates from the fact that  the 
 reflectionless system
 $H_n$ is related to the free particle  $H_0$
 by the chain of the subsequent Darboux transformations,
 and  the $H_0$ possesses a nontrivial integral
of motion  $p=-i\frac{d}{dx}$.

\section{Soliton scattering and Darboux transformations}

Besides the discussed pairs ($H_n$, $H_{n-1}$ )
of reflectionless Schr\"odinger 
systems
related by the first order Darboux intertwining operators, 
there is  another class of such systems,
for which the paired Hamiltonians are also interrelated by the first 
order Darboux generators. Unlike the described case,
the reflectionless partners in these pairs  are completely isospectral.
The corresponding  $n$-soliton partner  potentials
$U_n(x;\kappa_1,\tau_1,\ldots,
\kappa_n,\tau_n)$ and $U_n(x;\kappa_1,\tilde{\tau}_1,\ldots,
\kappa_n,\tilde{\tau}_n)$ are characterized by the same 
scaling parameters $\kappa_i$, $i=1,\ldots,n$,
but different sets of the translation parameters correlated as follows \cite{PAM+}\,:
\begin{equation}\label{Deltataun}
	\tau_i-\tilde{\tau}_i=
	\frac{1}{\kappa_i}\text{arctanh}\,\frac{\kappa_i}{\mathcal{C}}=
	\frac{1}{2\kappa_i}\log\frac{\mathcal{C}+\kappa_i}{
	 \mathcal{C}-\kappa_i}\,,
\end{equation}
where $\mathcal{C}$ is an additional  real parameter such that 
$\vert\mathcal{C}\vert>\kappa_n$.
A comparison of the quantities (\ref{Deltataun}) and  (\ref{phasen=2})
indicates that   (\ref{Deltataun}) can be related somehow 
to the effect of the scattering of solitons.
In this section we  show how
each indicated family of the paired reflectionless systems,
with partners intertwined by the first order Darboux generators,  
can be transformed  into another by a certain
limit procedure, which admits a  soliton scattering
interpretation.



To this aim, we  first consider  the limits $\tau_n\rightarrow \pm \infty$ 
applied to the  reflectionless system $H_n$.
To study the induced deformation of the potential
$U_n$  and  superpotential $\cW_n$ (the latter 
will play a role of the 
potential for an  associated Dirac system),
 it is sufficient to investigate the limits of the prepotential
 $\Omega_n$ because of the  relations  
$2\frac{d}{dx}\Omega_{n}=U_n$,  and $\Omega_{n}-\Omega_{n-1}=\cW_n$.
We shall demonstrate that 
$\Omega_{n}= -\frac{d}{dx}\log\W_n
\rightarrow \tilde\Omega_{n-1}(\cC)-\cC$ 
for  $\tau_n\rightarrow\pm\infty$, 
where  $\cC=\pm \kappa_n$, and $\tilde\Omega_{n-1}$ is identical to
$\Omega_{n-1}$ with $\tau_i$, $i=1,\ldots, n-1$,  changed for 
$\tilde{\tau}_i=\tau_i
-\f{1}{2\kappa_i}\log{\f{\cC+\kappa_i}{\cC-\kappa_i}}$.
{}From here it follows also that if we apply  subsequently another limit
 $\kappa_n\rightarrow \kappa_{n-1}$,
 or that is the same,  
  $\tilde\tau_{n-1}\rightarrow \mp\infty$, 
  the deformed (by $\kappa$-dependent  $\tau$-displacements) 
  prepotential transforms  as
   $ (\tilde\Omega_{n-1}-\cC)\rightarrow\Omega_{n-2}$.
So, the effect of sending subsequently 
the  two solitons with indices $n$ and $n-1$ 
to infinity in opposite directions
results in disappearing of the two bound states from the spectrum, 
without changing the rest of the $2(n-2)$ soliton parameters 
in the reflectionless potential  $U_{n-2}$. 
This corresponds to  a fermion-like behaviour of solitons 
already mentioned below Eq. (\ref{Deltaphin}).
\vskip0.1cm

In the limit $\tau_n\rightarrow\pm\infty$, for the
prepotential 
$\Omega_n=-(\log\W(\psi_1,\cdots,\psi_n))_x$ we find that 
$\Omega_n\rightarrow-(\log\W(\psi_1,\cdots,\psi_{n-1}, 
C_n^\pm e^{\pm\kappa_nx}))_x$, 
where  $C_n^\pm=\epsilon_n^\pm \frac{1}{2}e^{\pm\kappa_n\tau_n}$ 
is an exponentially  divergent multiplicative factor  with
$\epsilon_n^+=1$ and $\epsilon_n^-=(-1)^{n+1}$.
By the Wronskian properties,  we have
$\W(\psi_1,\cdots,\psi_{n-1}, C_n^\pm e^{\pm\kappa_nx})=
C_n^\pm \W(\psi_1,\cdots,\psi_{n-1},  e^{\pm\kappa_nx})$. 
The logarithmic derivative eliminates the  
$x$-independent divergent
multiplicative factor $C_n^\pm$, 
and in the limit  $\tau_n\rightarrow\pm\infty$ we obtain 
$\Omega_n\rightarrow-(\log\W(\psi_1,\cdots,\psi_{n-1}, 
e^{\pm\kappa_nx}))_x$. 
We note now that 
$\W(\psi_1,\cdots\psi_{n-1},e^{\pm\kappa_nx})=
e^{\pm\kappa_nx}\det  \|{W}_n^\natural\|$,  where 
\be\label{undW}
\|{W}_n^\natural\|=\left(%
    \begin{array}{ccccc}
     \ch \kappa_1x_1 &\sh \kappa_2x_2&\ldots&\psi_{n-1} & 1 \\
     \kappa_1\sh \kappa_1x_1&\kappa_2\ch \kappa_2x_2& 
		&\partial_x\psi_{n-1} &\pm\kappa_n\\
    \vdots& &\ddots&\vdots&\vdots\\
    \partial^{n-1}_x\ch \kappa_1x_1&\partial^{n-1}_x\sh \kappa_2x_2&
    \ldots&\partial^{n-1}_x\psi_{n-1}&(\pm1)^{n-1}\kappa_n^{n-1}
    \end{array}%
    \right),
\ee
and   $x_i\equiv x+\tau_i$. 
By changing the rows $L_j$, $j=1,\ldots,n-1$,  of the matrix (\ref{undW})
for the linear combinations\,: $L_j\rightarrow\kappa_n L_j\mp L_{j+1}$ ,
we find  that 
 $(\log\det \|{W}_n^\natural\|)_x=(\log\det\|\hat{W}_n\|)_x$
where 
\begin{equation}
    \|\hat{W}_n\|=
    \left(
\begin{array}{ccccc}
 {\rm Ch}_1^\mp &   {\rm Sh}_2^\mp & {\rm Ch}_3^\mp & \ldots &0  \\
 \kappa_1 {\rm Sh}_1^\mp  & \kappa_2{\rm Ch}_2^\mp &
  \kappa_3 {\rm Sh}_3^\mp & \ldots &0 \\
  \kappa_1^2{\rm Ch}_1^\mp &
  \kappa_2^2 {\rm Sh}_2^\mp & \kappa_3^2{\rm Ch}_3^\mp & \ldots &0  \\
 .&.&\ldots&.\\
  \partial_x^{n-1}\cosh\kappa_1x_1&\partial_x^{n-1}
  \sinh\kappa_2x_2& \partial_x^{n-1}\cosh\kappa_3x_3&
  \ldots&
  (\pm 1)^{n-1}\kappa_n^{n-1}
\end{array}
\right).
\end{equation}
Here we denote
  ${\rm Ch}_{i}^\mp=\kappa_n\cosh\kappa_{i} x_{i} \mp
\kappa_{i}\sinh \kappa_{i}x_{i}$ and  ${\rm
Sh}_{i}^\mp=\kappa_n\sinh\kappa_{i}x_{i} \mp \kappa_{i}\cosh
\kappa_{i}x_{i}$, ${i}=1,\ldots, n-1$, where the signs $-$ and $+$ 
correspond to the limits   $\tau_n\rightarrow
+\infty$ and  $\tau_n\rightarrow -\infty$, respectively.
These functions can be represented equivalently as 
 ${\rm
Ch}_{i}^\mp=\sqrt{\kappa_n^2-\kappa_{i}^2}
\cosh\kappa_{i}(x+\tau_{i}\mp\varphi_{i})$ and  ${\rm
Sh}_{i}^\mp=\sqrt{\kappa_n^2-\kappa_{i}^2}
\sinh\kappa_{i}(x+\tau_{i}\mp\varphi_{i})$, where
$\varphi_{i}=\frac{1}{2\kappa_{i}}\log
\frac{\kappa_n+\kappa_{i}}{\kappa_n-\kappa_{i}}$,
${i}=1,\ldots, n-1$.  As a consequence, we find that  
$(\log\W(\psi_1,\cdots,\psi_{n-1},e^{\pm\kappa_nx}))_x=
\pm\kappa_n+(\log\W(\tilde\psi_1,\cdots,\tilde\psi_{n-1}))_x$,
where  $\tilde\psi_i$ is identical to  $\psi_i$ but with $\tau_i$, $i=1,\ldots,n-1$,
 changed
for 
\begin{equation}\label{tildetaui}
	\tilde\tau_i=\tau_i\mp\frac{1}{2\kappa_{i}}\log
	\frac{\kappa_n+\kappa_{i}}{\kappa_n-\kappa_{i}}  \qquad
	\text{for}\quad \tau_n\rightarrow\pm\infty\,,
\end{equation}
that 
translates  finally into the transformation
$
\Omega_n\underset{\tau_n\rightarrow\pm\infty}
\longrightarrow\tilde\Omega_{n-1}\mp\kappa_n\,.
$
Note that $\tilde{\tau}_i-\tau_i$ given by (\ref{tildetaui})
corresponds to the change of the phase
in the two-soliton scattering given by the first relation in 
 (\ref{phasen=2}),  with indexes  $1$ and $2$  changed for $i$
 and $n$, respectively.

In the limit
 $\tau_n\rightarrow+\infty$ we find 
 that
\begin{equation}\label{t-inf}
	A_n=\frac{d}{dx}+\cW_n\,\,\rightarrow\,\,
	\frac{d}{dx}-\Delta_{n-1}(\kappa_n)
	=-X^\dag_{n-1}(\kappa_n)\,,
\end{equation}
where
\be\label{Xn-1}
	X_{n-1}=\frac{d}{dx}+\Delta_{n-1}\,,\qquad
	\Delta_{n-1}(\kappa_n)=
	\Omega_{n-1}-\tilde\Omega_{n-1}(\kappa_n)+\kappa_n\,.
\ee
The subsequent application of the  limit 
$\kappa_{n}\rightarrow \kappa_{n-1}$ gives 
\be\label{limX}
	X_{n-1}(\kappa_n)\rightarrow A_{n-1}\,,\qquad
	\tilde{A}_{n-1}\rightarrow X_{n-2}(\kappa_{n-1})\,,
\ee
where the first order operator $\tilde{A}_{n-1}$ is of the same form as $A_{n-1}$ but
with the parameters $\tau_i$ changed for
$\tilde{\tau}_i=\tau_i-\f{1}{2\kappa_i}
\log \f{\cC+\kappa_i}{\cC-\kappa_i}$.
The relations corresponding to the limit
$\tau_n\rightarrow-\infty$  can be written down explicitly  
in a similar way.

Since the $n$-soliton potentials are given by the relation
 $U_n=2\frac{d}{dx}\Omega_n$,  
by taking the limit  
 $\tau_n\rightarrow+\infty$ we eliminate 
 the bound state with $E_n=-\kappa_n^2$  from
  the spectrum of  $H_n$, and obtain 
 a new Hamiltonian with  $(n-1)$ bound states,  which we call 
 $\tilde{H}_{n-1}$. This hamiltonian is isospectral to  $H_{n-1}$ 
 but  each soliton in it is displaced with a phase dependent on 
$\kappa_n$, 
 \be\label{HTH}
	H_n(\tau_i)\underset{\tau_n\rightarrow +\infty}
	\longrightarrow H_{n-1}(\tilde{\tau}_i)
	\equiv\tilde{H}_{n-1}(\kappa_n)\,,\qquad
	\tilde{\tau}_i=\tau_i-\f{1}{2\kappa_i}
	\log \f{\kappa_n+\kappa_i}{\kappa_n-\kappa_i}\,.
\ee
The limit 
$\tau_n\rightarrow-\infty$  corresponds to the change 
of $\kappa_n$ for $-\kappa_n$  in (\ref{HTH}).
In general, from the viewpoint of  $\tilde H_{n-1}$,  
the $\kappa_n$ (or $-\kappa_n$)
is just an additional  parameter,  and from now on we call  
$\tilde H_{n-1}\equiv\tilde H_{n-1}(\cC)$, 
assuming for the sake of definiteness  that  $\cC>\kappa_{n-1}$.

On the other hand both the Hamiltonians
 $H_n$ in the limit $\kappa_n\rightarrow \kappa_{n-1}$  
 and  $\tilde{H}_{n-1}$ in the limit $\cC\rightarrow\kappa_{n-1}$ 
 correspond to a Hamiltonian  $H_{n-2}$,
 \be\label{HnCHn-1}
	H_n\underset{\kappa_n\rightarrow \kappa_{n-1}}
	\longrightarrow H_{n-2},\qquad\tilde{H}_{n-1}(\cC)
	\underset{\cC\rightarrow \kappa_{n-1}}
	\longrightarrow H_{n-2}\,.
\ee

As analogs of factorizations   (\ref{fact})
we obtain
\be\label{Xinter}
X_{n}^\dag X_{n}=\tilde{H}_{n}+\cC^2,\qquad
X_{n} X_{n}^\dag={H}_{n}+\cC^2\,,
\ee
where $X_n$ is defined in (\ref{Xn-1}) with index $n-1$ changed for $n$,
and  it is assumed here that $\cC^2>\kappa_n^2$.
In correspondence with (\ref{Xinter}), $X_n$ and  $X_n^\dagger$ not only factorize 
the isospectral Hamiltonians, but also intertwine them, $X_n\tilde{H}_n=H_nX_n$,
 $X_n^\dagger {H}_n=\tilde{H}_nX_n^\dagger$.
We also have the factorization relations 
\be
	\tilde{A}_{n} \tilde{A}^\dag_{n}=\tilde{H}_{n}+\kappa_{n}^2,\qquad
	\tilde{A}^\dag_{n} \tilde{A}_{n} =\tilde{H}_{n-1}+\kappa_{n}^2\,.
\ee
Using these last  relations, one 
can construct the generators which intertwine  
 $\tilde{H}_n$ and  $H_n$, being the  differential operators of the order $2n$,
\be\label{Ycaln}
	\mathcal{Y}_n=\A_n\tilde{\A}_n^\dag,
	\qquad \mathcal{Y}_n^\dag=\tilde{\A}_n\A_n^\dag\,,
\ee
$\mathcal{Y}_n\tilde{H}_n=H_n\mathcal{Y}_n$,
$\mathcal{Y}_n^\dagger H_n=\tilde{H}_n \mathcal{Y}_n^\dagger$,
where 
 $\tilde{\A}_n$ is defined as in  \ref{AnCD}  but with 
$A_i$ changed for $\tilde A_i$.

Another pair of important identities is
\be\label{XAAX}
	A_n X_{n-1}=X_n\tilde{A}_n\,,
	\qquad X_{n}^\dag A_n=
	\tilde{A}_n  X_{n-1}^\dag\,.
\ee
The operators appearing in the first equality
intertwine the Hamiltonians 
 $\tilde H_{n-1}$ and 
$H_n$, 
 $(A_n X_{n-1})\tilde H_{n-1}= H_n(A_n X_{n-1})$,
$(X_n\tilde{A}_n)\tilde H_{n-1}= H_n(X_n\tilde{A}_n)$,
and the equal operators from 
the other relation
intertwine in a similar manner 
$H_{n-1}$  and $\tilde{H}_n$.
The Hermitian conjugate form
of the operators from (\ref{XAAX}) 
intertwine the indicated pairs of the Hamiltonians 
 in the opposite direction.
The relations in  (\ref{XAAX})
are equivalent  to the identity
\begin{equation}\label{COnx}
    (\mathcal{C}+\Omega_{n-1}-\tilde{\Omega}_n)
    (\Omega_n-\tilde{\Omega}_n-\Omega_{n-1}+\tilde{\Omega}_{n-1})
    =
    (\tilde{\Omega}_{n}-\Omega_{n-1})_x\,,
\end{equation}
which, in turn,  is reduced to trigonometric identities
 \cite{PAM+}. 
 In the limit 
 $\tau_n\rightarrow \infty$, we find then that 
 the  
 intertwining between  $H_{n-1}$ and $H_n$ operator 
  $\mathcal{B}_n$, see
 Eq. (\ref{Bninter}),  transforms into  
 \begin{equation}\label{Bnlim}
 \mathcal{B}_n\rightarrow
 \left(\tilde{H}_{n-1}(\kappa_n)+\kappa_n^2\right)
 \mathcal{Y}_{n-1}^\dagger(\kappa_n) -2\kappa_n\left(\prod_{i=1}^{n-1}
 \left(\tilde{H}_{n-1}(\kappa_n)+\kappa_i^2\right)\right)X_{n-1}^\dagger(\kappa_n).
\end{equation}
This is a  reducible intertwining operator 
for a pair $H_{n-1}$ and $\tilde{H}_{n-1}$.
{}From  (\ref{Bnlim}) we extract the irreducible operators
$\mathcal{Y}_{n-1}^\dagger$ and $X_{n-1}^\dagger(\kappa_n)$
which intertwine the Hamiltonians 
 $H_{n-1}$  and $\tilde{H}_{n-1}$.
At the same time, for the Lax-Novikov integral  $\mathcal{L}_n$
we have
 \begin{equation}\label{Lnlim}
 \mathcal{L}_n\rightarrow
 \left(\tilde{H}_{n-1}(\kappa_n)+\kappa_n^2\right)
 \tilde{\mathcal{L}}_{n-1},
\end{equation}
that provides us with   the  irreducible 
nontrivial integral $ \tilde{\mathcal{L}}_{n-1}$  for $\tilde{H}_{n-1}$
\footnote{The questions of redundancy of 
nonlinear supersymmetric algebra in a general
context were studied in 
\cite{ASo}, see also the recent review \cite{AIo}.
The very nontrivial picture
of redundancy and transmutations appearing  in the completely isospectral 
supersymmetric pairs of reflectionless systems 
was investigated in detail in  \cite{PAM,PAM+}.}.
 
\begin{figure}[h!]\begin{center}
    \includegraphics[scale=0.6]{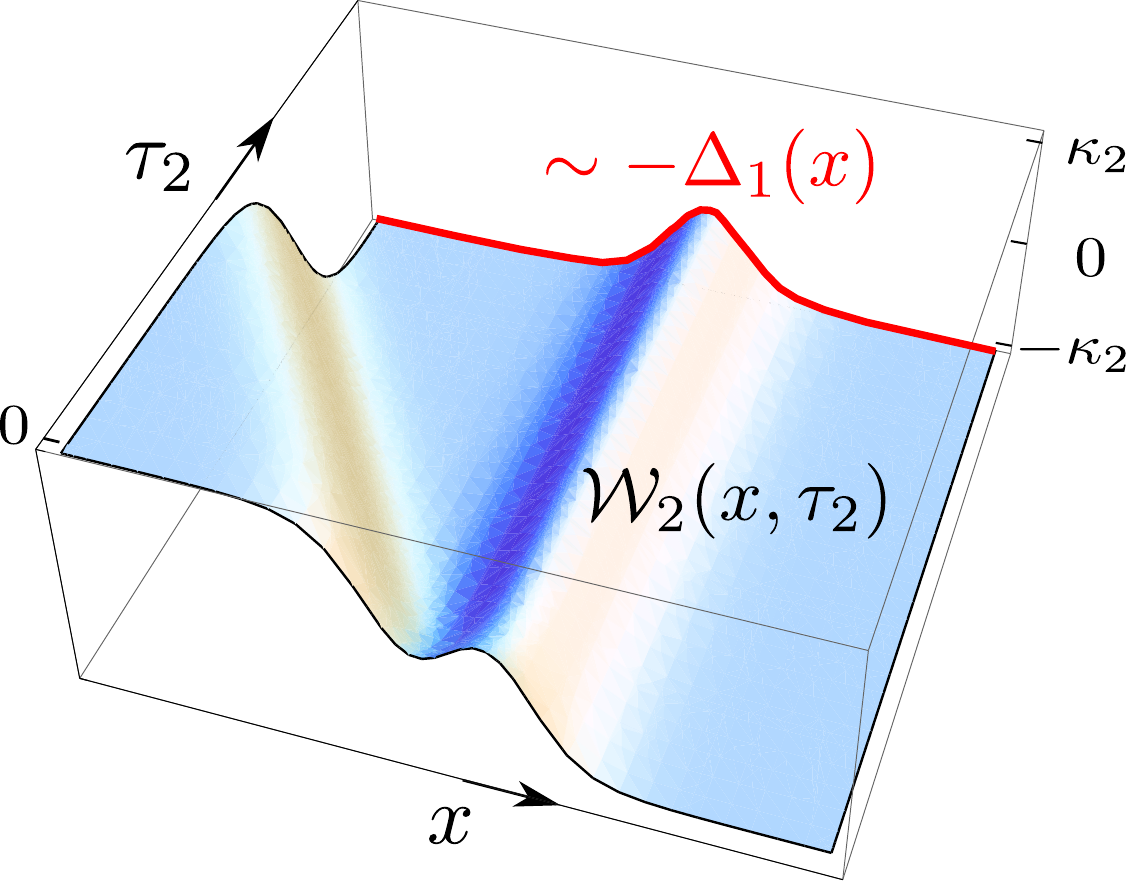}
    \caption{In the limit $\tau_n\rightarrow \infty$,  a
    topologically nontrivial superpotential $\mathcal{W}_n$ (being also the 
    corresponding 
    scalar Dirac potential) with 
    asymptotic behaviour 
    $\lim_{x\rightarrow -\infty}\mathcal{W}_n(x)=
    -\lim_{x\rightarrow +\infty}\mathcal{W}_n(x)=\kappa_n>0$      
        transforms  (asymptotically) into  a topologically trivial
        superpotential $-\Delta_{n-1}$ such that 
        $\lim_{x\rightarrow -\infty}\Delta_{n-1}(x)=
        \lim_{x\rightarrow +\infty}\Delta_{n-1}(x)=\kappa_n>0$. 
    This corresponds 
    to sending the $n$-th kink  to 
    $x=-\infty$. The figure corresponds to the case $n=2$, and 
    shows the superpotential $\mathcal{W}_2$ as a function of 
    $x$ and $\tau_2$.}
   \label{fig1}
\end{center}
\end{figure}

\begin{figure}[h!]\begin{center}
    \includegraphics[scale=0.6]{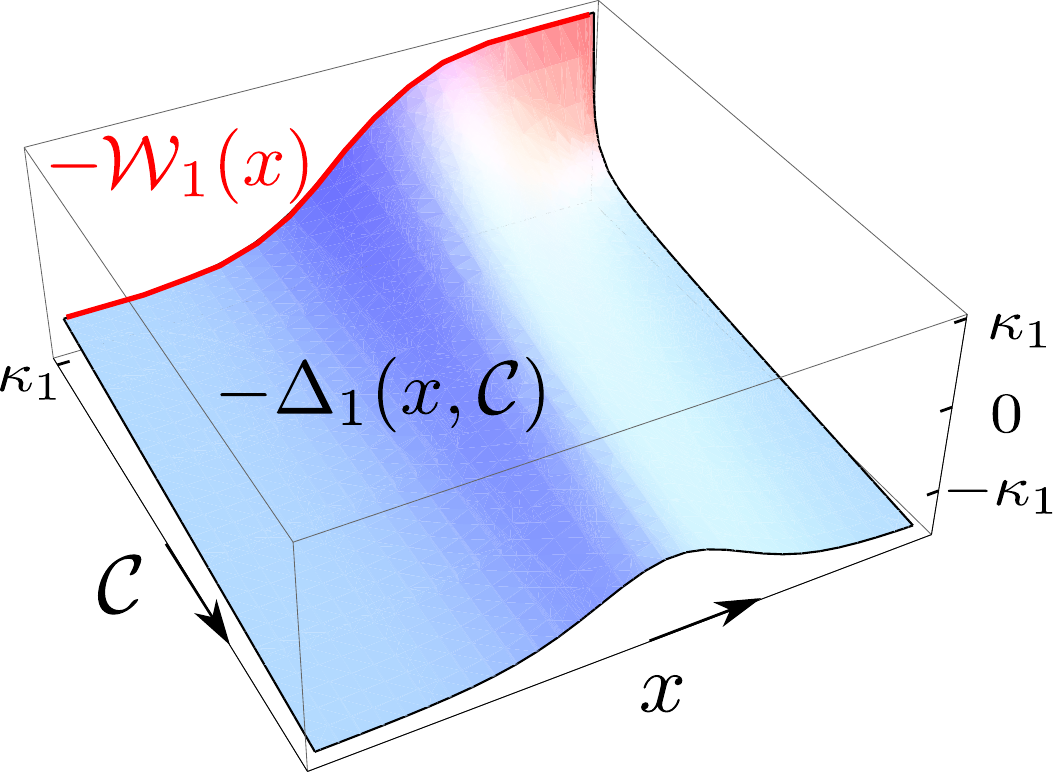}
    \caption{A topologically trivial  superpotential $\Delta_{n}$ 
    transforms into a topologically nontrivial 
     superpotential  $\mathcal{W}_n$ 
    through the limit $\vert \tilde{\tau}_n\vert \rightarrow\infty$, which 
    is equivalent to  the limit $\cC^2\rightarrow \kappa_{n}^2$.
    The figure illustrates the case when  
     the kink-antikink  Dirac potential  with $n=1$ 
    transforms  in the limit  $\cC\rightarrow\kappa_1$ into 
    the antikink potential.}
    \label{fig2}
\end{center}
\end{figure}

\begin{figure}[h!]\begin{center}
    \includegraphics[scale=0.6]{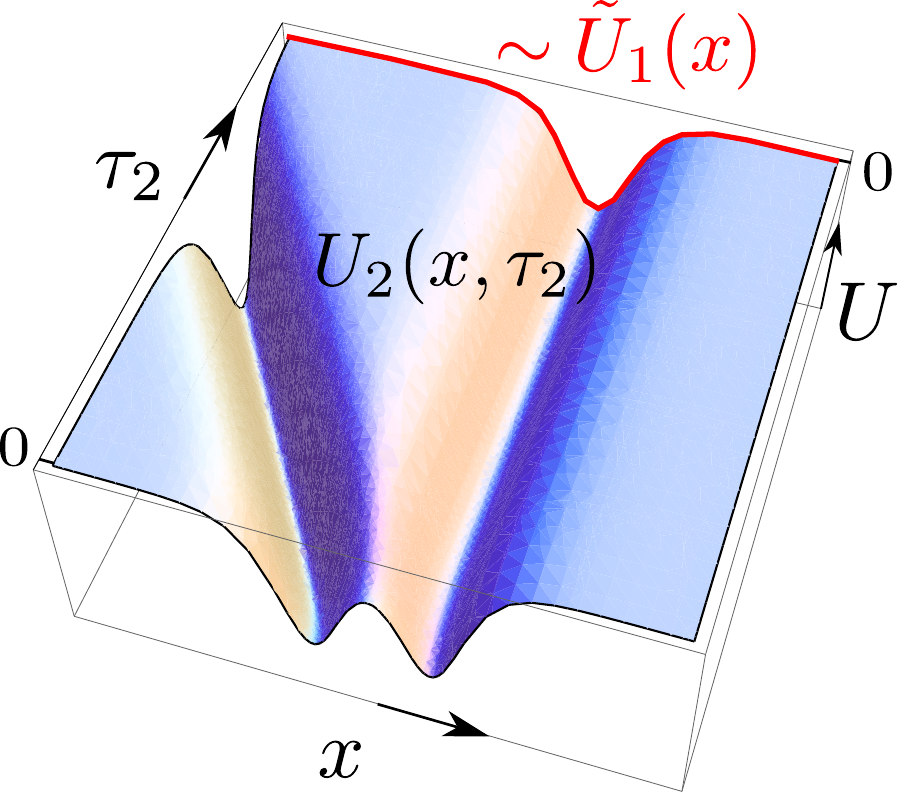}
    \caption{ For the particular case of $n=2$,  the figure illustrates 
    the transformation of the Schr\"odinger  $n$-soliton potential $U_n$
    into the $(n-1)$-soliton potential  $\tilde{U}_{n-1}$ in the limit 
    $\tau_n\rightarrow \infty$. }
    \label{fig3}
\end{center}
\end{figure}

\begin{figure}[h!]\begin{center}
    \includegraphics[scale=0.6]{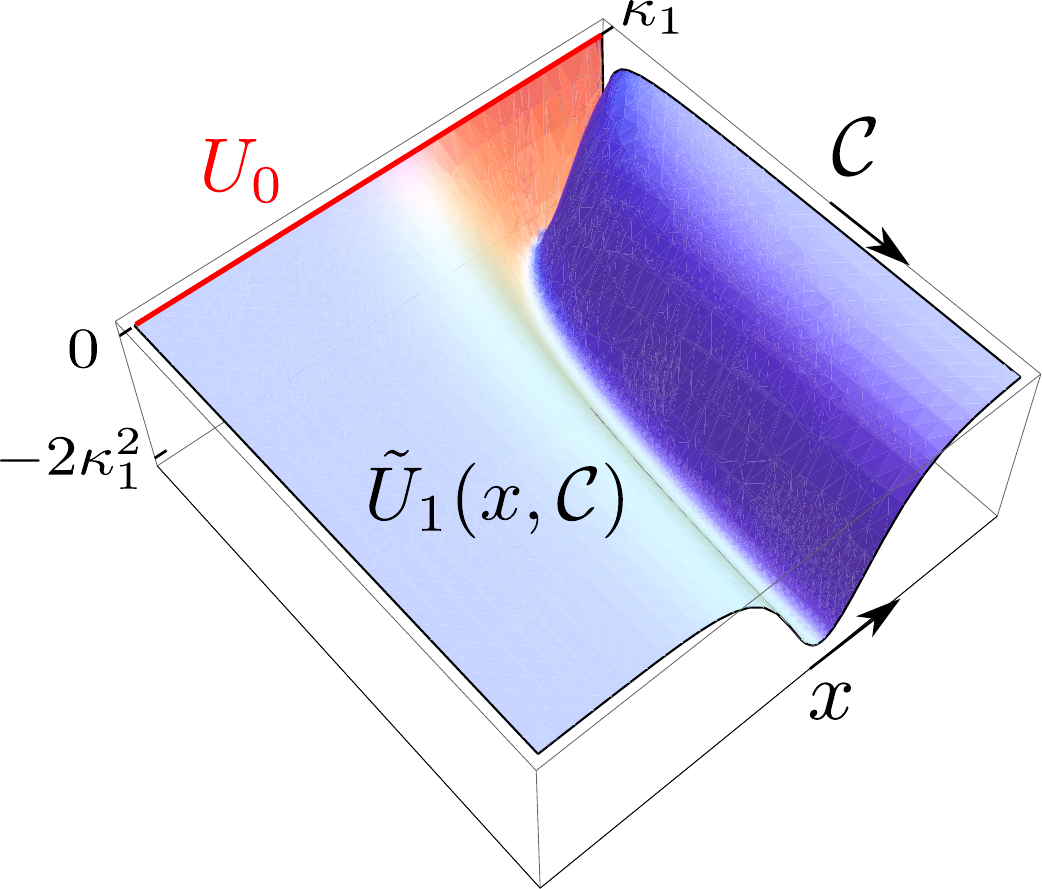}
    \caption{As an illustration for  the second limit in (\ref{HnCHn-1}),
    it is shown the transformation of  the one-soliton potential $\tilde{U}_1(x,\cC)$
    into the zero potential of the free particle case  in the limit $\cC\rightarrow\kappa_1$.
	Note that  in another  limit  $\cC\rightarrow\infty$, we have
		$\tilde{H}_{n} \rightarrow H_{n}$,  but
		the intertwining operator  $X_{n}$ blows up. Changing $X_n$ 
		for the rescaled 
		operator 
		$\hat{X}_{n}=\frac{1}{\cC}X_{n}$,  we get in the indicated limit 
		the trivial   operator,
		 $\hat{X}_n\rightarrow 1$,  as an intertwiner 
		 between  the two  identical copies of the reflectionless Schr\"odinger 
		 Hamiltonian $H_{n}$.}
    \label{fig4}
\end{center}
\end{figure}

  Figures \ref{fig1} and  \ref{fig2} illustrate different limits for superpotentials 
 $\mathcal{W}_n$ and $\Delta_n$, while figures \ref{fig3} and \ref{fig4} 
 show the transformations between potentials $U_n$ and $\tilde{U}_n$.
 
\vskip0.1cm

We have considered the limit when the translation parameter
$\tau_n$ in the $n$-soliton potential $U_n$
tends to infinity. 
It is interesting to  see what happens with reflectionless
system $H_n$ when we take the limit $\tau_j\rightarrow\pm\infty$ 
with $j<n$. 
Considering the same procedure as in the case $j=n$,
we find that the prepotential $\Omega_n$ changes 
for ${\Omega'}_{n-1}$, in which instead of (\ref{tildetaui})
the arguments $\tau_i$ are replaced by 
\begin{equation}
	\tau_i'=\left\{\begin{matrix}\tau_i\mp\frac{1}{2\kappa_i}\log \frac{\kappa_j+\kappa_i}{\kappa_j-\kappa_i}
	&\text{for}\,\,\, i<j\,,\\
	\tau_i\pm\frac{1}{2\kappa_i}(\log \frac{\kappa_j+\kappa_i}{\kappa_i-\kappa_j}
	+i\pi)
	&\text{for}\,\,\, i>j\,.\end{matrix}\right.
\end{equation}
For $i>j$ we have $\cosh\kappa_i(x+\tau'_i)=\pm i
\sinh\kappa_i(x+\hat{\tau}_i)$,
$\sinh\kappa_i(x+\tau'_i)=\pm i
\cosh\kappa_i(x+\hat{\tau}_i)$, 
where 
\begin{equation}\label{hattau}
	\hat{\tau}_i=\tau_i\mp\frac{1}{2\kappa_i}\log \bigg|
	\frac{\kappa_j+
	\kappa_i}{\kappa_j-\kappa_i}
	\bigg|\,.
\end{equation}
The effect of the limit $\tau_j\rightarrow \pm\infty$ results then  in the reduction
of the reflectionless system $H_n(x;\kappa_1,\tau_1,\ldots,\kappa_n,\tau_n)$ 
into the reflectionless system 
$\hat{H}_{n-1}$, where the latter Hamiltonian is given by the 
set of parameters $\kappa_i$ and $\hat{\tau}_i$ with $i=1,\ldots, j-1,j+1,\ldots,n$.
It is also  easy to check  that the application of the limit 
$\kappa_j\rightarrow \kappa_{j+1}$ with $j$ taking one of the values from the
set $1,\ldots, n-1$, transforms $H_n$ into $H_{n-2}$, where the latter reflectionless
Hamiltonian is characterized by the 
parameters $\kappa_i$ and $\tau_i$ with 
$i=1,\ldots, j-1,j+1,\ldots,n$. The same effect can be obtained if we apply subsequently two limits, 
first
$\tau_j\rightarrow +\infty$ (or $\tau_j\rightarrow -\infty$) and 
then $\hat{\tau}_{j-1}\rightarrow- \infty$ (or $\hat{\tau}_{j-1}\rightarrow+\infty$),
i.e. sent the soliton $j$ and the transformed one
with index $j-1$ to infinity in the opposite directions.

Note here that   applying appropriately the described limits
with $\tau_j$ tending to $+\infty$ or $-\infty$, 
we can reproduce exactly  the phases from (\ref{Deltaphin}), which
correspond to the soliton scattering picture  in the $n$-soliton solution 
for the KdV equation. 
Indeed, let us fix index $i=l$, where $1\leq l\leq n$.
For the sake of generality, assume that $1<l<n$.
Now, let us  take a limit  $\tau_n\rightarrow +\infty$. The  displaced 
value of $\tau_l$ will be given by the upper sign case of Eq. (\ref{hattau})  
with $i=l$ and $j=n$. Then we send subsequently to $+\infty$
the soliton indexed by   $j=n-1$, then $j=n-2$, etc., 
till $j=l+1$. Repeating analogous procedure with sending to $-\infty$ 
first the soliton with $j=1$, then with $j=2$, etc., till
$j=l-1$, the resulting changed translation parameter  will be given 
exactly by Eq. (\ref{Deltaphin}) corresponding to the case 
$t\rightarrow -\infty$ with $\tau^0_l$ changed for our
initial value  $\tau_l$. The sign  minus in the limit $t\rightarrow -\infty$ 
(in comparison with the sign in the limit $\tau_n\rightarrow +\infty$) is associated with 
the minus sign appearing in Eq. (\ref{Unxt}) before the term
$4\kappa_j^2t$.

\vskip0.2cm

Considering  the pairs of reflectionless Hamiltonians,
$(H_n,H_{n-1})$ or $(H_n,\tilde{H}_n)$, 
the partners of which are related by the first order Darboux intertwining 
generators,   we shall see below that the limits $\tau_n\rightarrow\pm \infty$ 
induce the transmutation of the type of the supersymmetry,
interchanging the cases of the unbroken  and broken supersymmetries.
On the other hand, the application of the limits $\tau_j\rightarrow\pm \infty$
with $j<n$ reduces only the number of the bound states in the partner Hamiltonians,
but does not change the type of the corresponding supersymmetry
of the extended reflectionless system.

The difference of the corresponding supersymmetries in the two cases 
 can be explained by  different nature of the first order Darboux 
 intertwining generators. In the case of the pairs $(H_n,H_{n-1})$,
 the intertwining generators $A_n$ and $A_n^\dagger$  
 are constructed in terms of the superpotential $\mathcal{W}_n$,
 see Eq. (\ref{An}), for which $\cW_n(x)\rightarrow\mp \kappa_n$ for
$x\rightarrow\pm\infty$.
This superpotential takes asymptotically the constant 
values of the opposite signs, and 
is topologically nontrivial. 
The Witten index for such extended system
 takes nonzero value,
	and the associated first order supersymmery (see the next section),
	is unbroken \cite{Witten,SUSYQM}.  The isospectral partners in the pairs 
	 $(H_n,\tilde{H}_n)$ are intertwined 
	by the first order Darboux generators $X_n$ and $X_n^\dagger$,
	constructed in terms of the superpotential $\Delta_n(\cC)$,
	see Eq. (\ref{Xn-1}) with $n-1$ changed for $n$.
	Since  $\lim_{x
	\rightarrow +\infty} \Delta_n=\lim_{x
	\rightarrow -\infty} \Delta_n=\cC$ with $\cC^2>\kappa_n^2>0$,
	the superpotential $\Delta_n(\cC)$ is topologically trivial,
	and the corresponding first order supersymmetry 
	will be broken in correspondence with zero value of
	the Witten index.

\section{Exotic supersymmetry of reflectionless 
systems with the first order supercharges}

Consider now an extended $2\times 2$ matrix Hamiltonian 
 $\cH={\rm diag}(H,H')$ with  $H$ and  $H'$ to be reflectionless systems,
 and identify $\Gamma=\sigma_3$ as a $\Z_2$-grading operator.
 As it was shown in   \cite{PAM}, in general case 
 such a  system is  characterized by exotic nonlinear supersymmetry 
 with two pairs of supercharges, which are the matrix higher order derivative 
 operators of the anti-diagonal form, constructed from the Darboux-Crum intertwiners.  
 The symmetry structure 
 of $\cH$ also has to include   two  higher order Lax-Novikov integrals
 of the subsystems $H$ and  $H'$. 
 Within this class of the extended reflectionless systems, there exist two special families,
 for which a pair of fermionic integrals 
 are the first order matrix differential operators
 of the form  $S_a=S_a^\dagger=\sigma_a\,
 {\rm diag}(D,D^\dag)$, $a=1,2$, which satisfy the relations
 $[S_a,\cH]=0$ and  $\{S_a,S_b\}=2\delta_{ab}(\cH+{\rm const})$.
 The operators $D$ and $D^\dagger$ 
 in this case not only intertwine the Hamiltonian operators 
$H$ and  $H'$, but also  factorize them,  
$H=D^\dag D+const$ and $H'=DD^\dag+const$ $\,$\footnote{
The supercharges, which are the higher order derivative operators,
factorize certain polynomials of the partner  Hamiltonians in correspondence with
relations of the form (\ref{BBHn}).}.

Without loss of generality one can choose 
$H=H_n$ to be reflectionless Hamiltonian with $n$-soliton potential.
Then there are only three possibilities to choose 
 $H'$ such that $H$ and $H'$ can be related  by the intertwining operators 
 of the first order.  These possibilities are $
 H'=H_{n-1}$, $H'= H_{n+1}$, or $H'=\tilde{H}_n(\cC)$.
 The trivial case of a  free particle, $H_0$, is exceptional\,:
 for it there are only two possibilities, $H'=H_1$ and $H'=H_0$,
 due to the translation invariance of $H_0$.
 
 Having this picture in mind, we first consider a class of the 
extended reflectionless ($2n+1)$-parametric systems
composed from isospectral  Hamiltonians each having $n$ bound states.
It is convenient to shift the Hamiltonian operators  for an additive 
constant term, and take
\begin{equation}\label{tHext+}
    \breve{\mathcal{H}}_{n}=\left(%
    \begin{array}{cc}
      H_n^\cC & 0 \\
      0 & \tilde H_{n} ^\cC\\
    \end{array}%
\right)
\end{equation}
as the extended Hamiltonian, where $H_n^\cC =H_n+\cC^2$,
$\tilde H_{n} ^\cC=\tilde{H}_{n} +\cC^2$.
A real constant  $\cC$ is   restricted here by the condition $\cC^2>\kappa_n^2$,
and $\tilde H_{n}=\tilde H_{n} (\cC)$ 
is the reflectionless system isospectral to
$H_n$ 
but with the parameters $\tau_i$ in the $n$-soliton potential
changed for 
the shifted set of translation parameters
$\tilde{\tau}_i$
given by equation (\ref{Deltataun}).
The spectra of the isospectral partner Hamiltonians are
$$
\sigma(H^\cC_n)=\sigma(\tilde{H}^\cC_n)=\cC^2-\kappa_n^2\cup\ldots
\cup \cC^2-\kappa_1^2\cup[\cC^2,\infty)\,.
$$
Each discrete energy level $\cC^2-\kappa_i^2$, $i=1,\ldots,n$, of the extended system 
(\ref{tHext+}), 
as well as the energy level $E=\cC^2$ at the edge of the continuous 
part of the spectrum are doubly degenerate.
At the same time,  each energy level inside the  conduction  band 
$(\cC^2,\infty)$
 of  $\breve{\mathcal{H}}_{n}$ is four-fold degenerate.
The set of the nontrivial integrals of motion (in addition to 
the trivial integral $\Gamma=\sigma_3$) 
of the supersymmetric system 
(\ref{tHext+}) consists of the two matrix differential operators of the first order 
composed from 
the Darboux intertwining generators of the form (\ref{Xn-1}) (with index 
$n-1$  changed for $n$),
\begin{equation}\label{tHext}
   \breve{\mathcal{S}}_{n,1}=\left(%
    \begin{array}{cc}
      0 & X_{n} \\
      X^\dag_{n} & 0 \\
    \end{array}%
\right),\qquad \breve{\mathcal{S}}_{n,2}=
i\sigma_3\breve{\mathcal{S}}_{n,1}\,.
\end{equation}
We have also two matrix integrals to be differential 
operators of the order $2n$ constructed 
from the intertwines (\ref{Ycaln}),
\begin{equation}\label{Qintn}
\breve{\mathcal{Q}}_{n,1}=\left(%
    \begin{array}{cc}
      0 & \mathcal{Y}_{n} \\
      \mathcal{Y}^\dag_{n} & 0 \\
    \end{array}%
\right),\qquad \breve{\mathcal{Q}}_{n,2}=
i\sigma_3\breve{\mathcal{Q}}_{n,1}\,.
\end{equation}
In addition, the system is characterized by the two diagonal matrix integrals 
constructed from the Lax-Novikov integrals (\ref{Pndef})
 of the subsystems, which are the differential operators 
of the order $2n+1$,
\begin{equation}
 \breve{\mathcal{P}}_{n,1}=\left(%
    \begin{array}{cc}
      \mathcal{L}_n & 0 \\
      0 & \tilde{\mathcal{L}}_n \\
    \end{array}%
\right),\qquad \breve{\mathcal{P}}_{n,2}=
\sigma_3\breve{\mathcal{P}}_{n,1}\,.\label{Integ}
\end{equation}
With the chosen $\Z_2$-grading operator $\Gamma=\sigma_3$, 
operators (\ref{tHext}) and (\ref{Qintn}) are identified as
the fermionic integrals, and  (\ref{Integ}) are identified as 
the bosonic 
generators. They, together with the Hamiltonian $\breve{\mathcal{H}}_n$, 
generate  the exotic superalgebra, whose 
nonzero (anti)-commutation relations are given by
\begin{equation}\label{SQsuper}
    \{\breve{\mathcal{S}}_{a},
    \breve{\mathcal{S}}_{b}\}=2\delta_{ab}\breve{\mathcal{H}},\qquad
    \{\breve{\mathcal{Q}}_{a},\breve{\mathcal{Q}}_{b}\}=
    2\delta_{ab}\breve{\P} ^2\,,\qquad
    \{\breve{\mathcal{S}}_{a},\breve{\mathcal{Q}}_{b}\}=
    2\delta_{ab}\mathcal{C}\breve{\P}
    +2\epsilon_{ab}\breve{\mathcal{P}}_{1}\,,
\end{equation}
\begin{equation}\label{P2rot}
    [\breve{\mathcal{P}}_{2},\breve{\mathcal{S}}_{a}]=
    2i\left(\breve{\mathcal{H}} 
    \breve{\mathcal{Q}}_{a}
    -\mathcal{C}\breve{\P} \breve{\mathcal{S}}_{a}\right)\,,\qquad
    [\breve{\mathcal{P}}_{2},\breve{\mathcal{Q}}_{a}]=
    2i\breve{\P} \left(\mathcal{C}\breve{\mathcal{Q}}_{a}
    -\breve{\P} \breve{\mathcal{S}}_{a}\right)\,,
\end{equation}
where $\breve{\P}_n=\prod_{j=1}^n(\breve{\mathcal{H}}_n
-\cC^2+\kappa_j^2)$,
 and  to simplify the expressions we omitted the index $n$
 in (\ref{SQsuper}), (\ref{P2rot}).
 Though our construction 
 with the two Schr\"odinger subsystems $H_n^\cC$ and $\tilde{H}_n^\cC$
 corresponds to the usual $\mathcal{N}=2$ supersymmetry generated by 
 the two supercharges $\breve{\mathcal{S}}_{n,a}$  to be matrix
 differential operators of the first order,
 we have obtained  the exotic supersymmetric structure
 with the two additional supercharges $\breve{\mathcal{Q}}_{n,a}$ 
 to be the higher order differential operators. In addition, 
 being the extended reflectionless  system,  it 
 also possesses two bosonic integrals of motion.
 The peculiarity of the present exotic supersymmetric structure 
 is that  the bosonic integral  $\breve{\mathcal{P}}_{n,1}$
 commutes with all the other integrals of motion, and
 plays a role of the central charge operator of the nonlinear 
 superalgebra\,\footnote{This is not so  in a general case of the 
 extended system composed from the two $n$-soliton  
 Schr\"odinger subsystems, see ref. \cite{PAM}.}.
 Another bosonic integral  $\breve{\mathcal{P}}_{n,2}$
realizes a rotation of the pairs of  the supercharges
$\breve{\mathcal{S}}_{n,a}$ and $\breve{\mathcal{Q}}_{n,a}$
by means of the commutation relations  (\ref{P2rot})
with the Hamiltonian-dependent structure coefficients.

Since the doublet of the ground states of 
$\breve{\mathcal{H}}_n$
has positive energy $\cC^2-\kappa_n^2>0$,
the first order supercharges $\breve{\mathcal{S}}_{n,a}$ 
do not annihilate them either, and the $\mathcal{N}=2$ Lie sub-superalgebra
generated by $\breve{\mathcal{S}}_{n,a}$  and 
$\breve{\mathcal{H}}_n$
corresponds to  the phase of the \emph{broken} supersymmetry. 
At the same time, according to Eq. (\ref{LHspectral}),
the doublet of the ground states is annihilated by the bosonic 
integrals $\breve{\mathcal{P}}_{n,a}$. Due to the second relation 
from  (\ref{SQsuper}), they are also annihilated by the 
higher order supercharges $\breve{\mathcal{Q}}_{n,a}$.
One can conclude therefore  that  the obtained exotic nonlinear 
$\mathcal{N}=4$ 
supersymmetry 
of the extended reflectionless system $\breve{\mathcal{H}}_n$
is \emph{partially broken}.\vskip0.1cm

Let us apply  now the limit $\cC^2\rightarrow\kappa_n^2$, 
associated  with the soliton scattering,
to the system  $\breve{\mathcal{H}}_n$.
For the sake of definiteness, let us assume that $\cC$ is positive,
and consider the limit $\cC\rightarrow \kappa_n$,
which corresponds to the limit $\tilde{\tau}_n\rightarrow-\infty$
for the subsystem $\tilde{H}_n^\cC$. In this limit, the Hamiltonian 
(\ref{tHext+}) and integrals of motion are transformed into
\begin{equation}\label{tHextnn}
     {\mathcal{H}}_{n}=\left(%
    \begin{array}{cc}
      H_n^{\kappa_n} & 0 \\
      0 &   H_{n-1}^{\kappa_n} \\
    \end{array}%
\right),\qquad  {\mathcal{S}}_{n,1}=\left(%
     \begin{array}{cc}
      0 & A_{n} \\
      A_{n}^\dag & 0 \\
    \end{array}%
\right),
\end{equation}
\begin{equation}\label{QP1kap}
 {\mathcal{Q}}_{n,1}=\left(%
    \begin{array}{cc}
      0 & \mathcal{B}_{n} \\
      \mathcal{B}^\dag_{n} & 0 \\
    \end{array}%
   \right),\qquad 
 {\mathcal{P}}_{n,1}=\left(%
   \begin{array}{cc}
      \mathcal{L}_{n} & 0 \\
      0 &H_{n-1}^{\kappa_n}\, \mathcal{L}_{n-1} \\
    \end{array}%
\right),
\end{equation}
and  the integrals with index $a=2$ are obtained  by the
same rule as in (\ref{tHext}), 
(\ref{Qintn}) and (\ref{Integ}),  
where the notations $H_n^{\kappa_n}=H_n+\kappa_n^2$
and
 $H_{n-1}^{\kappa_n}=H_{n-1}+\kappa_n^2$ are used.
To obtain the limit we have taken into account 
the relations (\ref{limX}), (\ref{Ycaln}) and  
(\ref{B2n}).  
The Hamiltonian $ {\mathcal{H}}_{n}$
and its integrals of motion 
generate the nonlinear superalgebra of the form similar to
(\ref{SQsuper}) and (\ref{P2rot}),
but with corresponding changes of the operators
on the right hand sides, and with the 
$\cC$ changed for $\kappa_n$.

Note that the lower matrix element 
in the integral  ${\mathcal{P}}_{n,1}$ (and similarly,
in  ${\mathcal{P}}_{n,2}$) is factorized
into the subsystem's Hamiltonian $H_{n-1}^{\kappa_n}$ and 
the corresponding Lax-Novikov integral.
The multiplicative factor $H_{n-1}^{\kappa_n}$ could be omitted there
without changing the property of commutativity 
of the diagonal matrix operators with the 
Hamiltonian  $ {\mathcal{H}}_{n}$.
However, this would change the property that 
the upper and lower matrix elements 
in the  integrals ${\mathcal{P}}_{n,a}$ 
are the differential operators of the same order $2n+1$, and,
as a consequence, 
would complicate the form of the superalgebra.

In spite of a similar  form of the superalgebra 
(with $\cC$ changed for $\kappa_n$),
the superextended  system we have here is essentially 
different from the previous one.
Indeed,  the system
 $ {\mathcal{H}}_{n}$, unlike  
 the  $\breve{\mathcal{H}}_{n}$, 
possesses now the non-degenerate ground state 
of zero energy, which corresponds to the lowest 
bound state of the upper subsystem 
$ {H}_{n}^{\kappa_n}$.  This state
is annihilated by all the four supercharges and 
the two bosonic integrals, and
the exotic nonlinear supersymmetry
we have here corresponds to the\emph{ unbroken}  phase.
Therefore, the limit we considered provokes the transmutation
of the partially broken exotic supersymmetry into the unbroken
 one.

Also, there exists a limit, associated with the soliton scattering,
which transmutes the exotic nonlinear supersymmetry 
from the unbroken phase into the partially 
broken exotic supersymmetry.  
To see this, we  apply to the system 
(\ref{tHext}), (\ref{QP1kap}) the limit $\tau_n\rightarrow\infty$, 
which corresponds to sending the soliton 
with index $n$ in the subsystem $H^{\kappa_n}_n$ to 
infinity.
We find then 
with the help of (\ref{Bnlim}) and (\ref{Lnlim})
that
\be\label{HSDiam}
	\mathcal{H}_{n}
	\underset{\tau_n\rightarrow\infty}
	\longrightarrow\breve{\mathcal{H}}^\Diamond_{n-1}\,,
	\qquad\mathcal{S}_{n,a}
	\underset{\tau_n\rightarrow\infty}
	\longrightarrow
	\breve{\mathcal{S}}^\Diamond_{n-1,a}\,,
\ee
\be\label{PQDiam}
	\mathcal{P}_{n,a}
	\underset{\tau_n\rightarrow\infty}
	\longrightarrow
	\breve{\mathcal{H}}^\Diamond_{n-1}\,\breve{\mathcal{P}}^\Diamond_{n-1,a}\,,
	\qquad\mathcal{Q}_{n,a}
	\underset{\tau_n\rightarrow\infty}
	\longrightarrow
	-\breve{\mathcal{H}}^\Diamond_{n-1}\,
	\breve{\mathcal{Q}}^\Diamond_{n-1,a}
	+2\kappa_n\breve{\P}^\Diamond_{n-1}\,
    \breve{\mathcal{S}}^\Diamond_{n-1,a}\,.
\ee
Here we have used the notation 
${F}^\Diamond=\sigma_2 {F}\sigma_2$, 
 which corresponds to a unitary transformation between the matrix 
 operators
 $$
F= \left(%
    \begin{array}{cc}
      \alpha & \beta \\
      \gamma & \delta \\
    \end{array}%
   \right)\qquad
   \text{and}\qquad 
  F^\Diamond= \left(%
    \begin{array}{cc}
      \delta & -\gamma \\
      -\beta & \alpha \\
    \end{array}%
   \right),
  $$
 and imply that   the operators indexed by $n-1$ 
 are given by the same expressions as the 
operators  associated with  $\breve{\mathcal{H}}_n$, but with 
 the parameter $\cC$ changed in the structure of the latter 
 operators
for   $\cC=\kappa_n$.  
  As a consequence,  we also obtain a four-term chain of 
  the limits
\be
\mathcal{H}_{n}
\underset{\tau_n\rightarrow\infty}
\longrightarrow\breve{\mathcal{H}}^\Diamond_{n-1}
\underset{\kappa_{n}\rightarrow \kappa_{n-1}}
\longrightarrow\mathcal{H}^\Diamond_{n-1}
\underset{\tau_{n-1}\rightarrow\infty}
\longrightarrow\breve{\mathcal{H}}_{n-2}\,.
\ee
Note that the multiplicative factor 
$\breve{\mathcal{H}}^\Diamond_{n-1}$ 
in the limit of  the operators 
$\mathcal{P}_{n,a}$ and $\mathcal{Q}_{n,a}$ in (\ref{PQDiam})
corresponds to a reduction of the order 
of the integrals that is related with the loss of the one eigenvalue
of zero energy in comparison with 
the spectrum of the system $\mathcal{H}_n$.

\section{Transparent Dirac systems}

We have discussed the Darboux-Crum  transformations, 
 the exotic supersymmetric structure  based on them, 
 and transmutations 
 of supersymmetry   in
 the reflectionless  systems described by the $2\times  2$ matrix 
 second order Scr\"odinger
Hamiltonian operators. 
One can take one of the two first order 
Hermitian supercharges
appearing in these second order systems, 
and consider it as a first order matrix Hamiltonian
for the $(1+1)$-dimensional Dirac system. 
We can identify then the Darboux--Crum
generators, which  intertwine such
reflectionless first order  matrix Hamiltonians.
This opens a possibility  to investigate  
exotic supersymmetry and its
transmutations in the transparent Dirac systems.

Let us take the first order supercharge $\breve{\mathcal{S}}_{n,1}$
from  (\ref{tHext}),  and identify it as the Dirac Hamiltonian,
  $\breve{H}^D_{n}\equiv\breve{\mathcal{S}}_{n,1}$.
 This system corresponds 
  to  the (1+1)-dimensional  Dirac particle
 in a scalar potential
 $\Delta_n(x)=\Omega_n-\tilde{\Omega}_{n}+\cC$ with asymptotic behaviour 
 $\Delta_n(x)\rightarrow \cC$ for 
$x\rightarrow\pm\infty$. 
Due to the relation  
of commutativity
$[\breve{\mathcal{S}}_{n,1},\breve{\mathcal{P}}_{n,1}]=0$,
the potentials of this form correspond
to the solutions of the 
multi-kink-antikink type for the stationary  mKdV 
hierarchy 
\cite{PAM+}.

The Dirac Hamiltonian $\breve{H}^D_n$ 
has $2n$ bound states, and its spectrum is  symmetric, 
\be
	\sigma(\breve{H}^D_{n})=(-\infty,-\cC]\cup \breve{\mathcal{E}}^-_1
		\cup\ldots\cup \breve{\mathcal{E}}^-_n
	\cup \breve{\mathcal{E}}^+_n\cup\ldots \cup 
	\breve{\mathcal{E}}^+_1\cup
	[\cC,\infty)\,,
\ee
where $\breve{\mathcal{E}}^\pm_i=\pm\sqrt{\cC^2- \kappa^2_{i}}$, $i=1,\ldots, n$,
and semi-infinite intervals $[\cC,\infty)$ and $(-\infty,-\cC]$
correspond to the doubly degenerate continuous parts of the spectrum.
In the limit  $\cC\rightarrow\kappa_n$,  we have 
$\breve{H}^D_{n}\rightarrow H^D_{n}=\mathcal{S}_{n,1}$, 
where $\mathcal{S}_{n,1}$ is defined in (\ref{tHextnn}).
A scalar potential takes here the form 
 $\cW_n(x)=\Omega_n-\Omega_{n-1}$,
with $\cW_n(x)\rightarrow\mp \kappa_n$ for
$x\rightarrow\pm\infty$.
The potentials of this form are, again, 
 the solutions of the kink (or, antikink) type
for the  stationary  mKdV  hierarchy
due to the relation   $[\mathcal{S}_{n,1},\mathcal{P}_{n,1}]=0$.
The spectrum of the  Dirac Hamiltonian $H^D_{n}$  has
 $2n-1$ bound states, including  one bound state of zero energy,
\be
	\sigma({H}^D_{n})=(-\infty,-\kappa_n]\cup \mathcal{E}^-_1
		\cup\ldots\cup \mathcal{E}^-_{n-1}\cup 0
	\cup \mathcal{E}^+_{n-1}\cup\ldots \cup \mathcal{E}^+_1\cup
	[\kappa_n,\infty)\,,
\ee
where ${\mathcal{E}}^\pm_i=\pm\sqrt{\kappa_n^2- \kappa^2_{i}}$, $i=1,\ldots, n-1$.
The  two  discrete energy levels 
$\breve{\mathcal{E}}^-_n$ and $\breve{\mathcal{E}}^+_n$ 
of the system $\breve{H}^D_n$ merge in the limit  $\cC\rightarrow\kappa_n$
and transform
into a non-degenerate zero energy level of the bound state for the system 
${H}^D_n$.

\subsection{First order matrix Darboux  intertwiners  for Dirac systems }

Let us return to the identity (\ref{XAAX}),
\be\label{axa}
	A_n(x,\tau_i) X_{n-1}(x,\tau_i,\cC)=X_n(x,\tau_i,\cC)A_n(x,\tilde{\tau}_i)\,,
\ee
where $X_n(x,\tau_i,\cC)=\frac{d}{dx}+\Delta_n(x,\tau_i,\cC)$, and
\be
	\Delta_n(x,\tau_i,\cC)=\Omega_{n}(x,\tau_i)-\Omega_{n}(x,\tilde{\tau}_i)+\cC,\quad
	\tilde{\tau}_i=\tau_i-\varphi_i(\cC)\,,\quad
	 \varphi_i(\cC)=\f{1}{2\kappa_i}\log{\f{\cC+\kappa_i}{\cC-\kappa_i}}\,.
\ee
If in (\ref{axa}) we change   $\tau_i\rightarrow\tau_i+\varphi_i(\cC)$,
then make a replacement  $\cC\rightarrow-\cC$,  and 
take into account that  $\varphi_i(-\cC)=-\varphi_i(\cC)$ and  that $X_n$ satisfies 
the relation $X_n(x,\tau_i-\varphi_i(\cC),-\cC)=-X^\dag_n(x,\tau_i,\cC)$, we 
obtain 
the identity
\be\label{axda}
	A_n(x,\tilde{\tau}_i) X^\dag_{n-1}(x,\tau_i,\cC)
	=X^\dag_n(x,\tau_i,\cC)A_n(x,\tau_i)\,.
\ee
Using the notations  $A_n\equiv A_n(x,\tau_i)$,
$\tilde A_n(\cC)\equiv A_n(\tilde{\tau}_i)$
and  $X_n(\cC) \equiv X_n(x,\tau_i,\cC)$, 
the equations 
(\ref{axa})--(\ref{axda}) and their Hermitian conjugate give us 
the relations 
\be\label{ax=xa}
	A_nX_{n-1}=X_n\tilde A_n\,,\qquad A^\dag_nX_{n}=
	X_{n-1}\tilde{A}_n^\dag\,,
\ee
\be
	\tilde{A}_nX^\dag_{n-1}=X^\dag_{n}A_n\,,
	\qquad\tilde{A}^\dag_nX^\dag_n=
	X^\dag_{n-1}A_n^\dag\,.
\ee
Using these relations,
we can define the intertwining  operator between
the Dirac Hamiltonians $\breve{H}^D_n$ and  
$\breve{H}^D_{n-1}$, which also is the intertwining operator between
the extended (supersymmetric) Schr\"odinger Hamiltonians
$\breve\cH_n$ and $\breve\cH_{n-1}$,
\begin{equation}\label{Acal1}
\breve{\mathcal{A}}_n=
    \left(%
    \begin{array}{cc}
    {A}_n & 0 \\
      0 & \tilde{A}_n \\
   \end{array}%
    \right)
	,\qquad
	\breve{\mathcal{A}}_n\breve{H}^D_{n-1}=
	\breve{H}^D_n\breve{\mathcal{A}}_n\,,\qquad 
	\breve{\mathcal{A}}_n\breve{\cH}_{n-1}=
	\breve{\cH}_n\breve{\mathcal{A}}_n\,.
\end{equation}
In  the limit
$\cC\rightarrow\kappa_n$, the relations in   (\ref{ax=xa}) 
are transformed into the trivial identity $	A_nX_{n-1}(\kappa_n)=A_nX_{n-1}(\kappa_n)$,
and the relation 
\begin{equation}
	A^\dag_nA_n=X_{n-1}(\kappa_n)X_{n-1}^\dag(\kappa_n)
	=H_{n-1}+\kappa_n^2\,,
\end{equation}
where we have used the limits (\ref{limX}).
These identities allow us to construct a new operator of
intertwining between the Dirac systems 
$H^D_n$ and  $\breve{H}^D_{n-1}$, 
and  between the super-extended 
Schr\"odinger Hamiltonians
$\cH_n$ and $\breve\cH_{n-1}$,  
\begin{equation}\label{Acal2}
\mathcal{A}_n=
    \left(%
    \begin{array}{cc}
      {A}_n& 0 \\
      0 & X_{n-1}(\kappa_n)\\
    \end{array}%
    \right),\qquad
\mathcal{A}_n\breve{H}^D_{n-1}(\kappa_n)=H^D_n\mathcal{A}_n\,,\qquad
\mathcal{A}_n\breve{\cH}_{n-1}(\kappa_n)=\cH_n\mathcal{A}_n\,,
\end{equation}
where we indicated a dependence of the corresponding operators on 
$\kappa_n=\cC$.

This construction corresponds here 
to  the Darboux transformations for reflectionless Dirac systems,
and, particularly, gives us a possibility 
to construct analytically the states of 
$H^D_n$ and $\breve H^D_n$ 
in terms of the eigenstates 
{$\breve\Phi_0$}  
of the  matrix operator 
$\breve H^D_0=-\sigma_2p+\sigma_1\cC$,
\be\label{H0DC}
\breve{H}^D_0(\cC)=\left(%
    \begin{array}{cc}
     0 & \frac{d}{dx}+\cC \\
      -\frac{d}{dx}+\cC &0 \\
    \end{array}%
    \right),
\ee
which corresponds to the Hamiltonian of the free massive Dirac particle. 
The eigenstates $\breve{\Phi}_n$  of  $\breve{H}^D_n$ 
can be presented in the form  
$
	\breve\Phi_n=\breve{\mathcal{A}}_{n}
	\breve{\mathcal{A}}_{n-1}\dots
	\breve{\mathcal{A}}_{1}\breve\Phi_0,
$
while  the eigenstates of  $H^D_n$ 
are constructed  in the form 
$
	\Phi_n=\mathcal{A}_{n}
	\breve{\mathcal{A}}_{n-1}\breve{\mathcal{A}}_{n-2}
	\dots\breve{\mathcal{A}}_{1}\breve{\Phi}_0\
$
in terms of the eigenstates 
 $\breve{\Phi}_0$ of  the Dirac Hamiltonian
$\breve H^D_0(\kappa_n)=-\sigma_2p+\sigma_1\kappa_n$.
The explicit form of the scattering states and  $2n$ 
bound states of  the $\breve H^D_n$ are given by 
\be\label{Phieps}
\breve\Phi^{\epsilon}_n\left(\breve{\mathcal{E}}^\pm(k^2)\right)=\left(%
    \begin{array}{cc}{\Psi}^\epsilon_n(k^2)\\
     \pm\sqrt{\f{\cC-i\epsilon k}{\cC+i\epsilon k}}
		\tilde\Psi^\epsilon_n(k^2)\\
    \end{array}%
    \right)
    ,\quad
\breve\Phi_n\left(\breve{\mathcal{E}}^\pm_i\right)=\left(%
    \begin{array}{cc}{\Psi}_n(-\kappa^2_i)\\
     \pm\tilde\Psi_n(-\kappa^2_i)\\
    \end{array}%
    \right)
    ,
\ee
where 
 $\breve{H}^D_n\breve\Phi_n(\breve{\mathcal{E}})
 =\breve{\mathcal{E}}\breve\Phi_n(\breve{\mathcal{E}})$,
$ \breve{\mathcal{E}}^\pm(k^2)=\pm\sqrt{\cC^2+k^2}$, 
$\breve{\mathcal{E}}^\pm_i=\pm\sqrt{\cC^2-\kappa_i^2}$, 
$i=1,\ldots,n$,
 $\Psi_n$ are Schr\"odinger eigenstates defined in  
 (\ref{est}),  and the parameter  $\epsilon=\pm 1$
corresponds to the two possible directions 
in which the waves can propagate.
 The two
 discrete energy levels  
$\breve{\mathcal{E}}_n^\pm=\pm\sqrt{\cC^2-\kappa^2_n}$  merge
 in the limit $\cC\rightarrow\kappa_n$,  
 and two corresponding eigenstates of $\breve{H}^D_n$
 reduce  to the unique state of zero energy
of the Dirac Hamiltonian $H^D_n$, 
\be \label{Phi0E0}
\breve\Phi_n\left(\breve{\mathcal{E}}_n^\pm\right)=\left(%
    \begin{array}{cc}{\Psi}_n(-\kappa^2_n)\\
     \pm\tilde\Psi_n(-\kappa^2_n)\\
    \end{array}%
    \right)\,\, \rightarrow\,\,\Phi_n\left(0\right)=\left(%
    \begin{array}{cc}{\Psi}_n(-\kappa^2_n)\\
     0\\
    \end{array}%
    \right) \,.
\ee

\subsection{Exotic supersymmetry of reflectionless Dirac systems}

The matrix operator
$\breve{\mathcal{P}}_{n,1}$
and the  Dirac Hamiltonian 
 $\breve{\mathcal{H}}^D_n$ correspond to the Lax pair 
 for  the $n$-th member of the stationary mKdV hierarchy, and the
  scalar Dirac potential
 $\Delta_n(x)$ is identified  as
 the corresponding soliton 
 (multi-kink-antikink type) solution.  
 Since  $[\breve{\mathcal{P}}_{n,1}, \breve{\mathcal{H}}^D_n ]=0$,
 the 
 $\breve{\mathcal{P}}_{n,1}$ is a nontrivial integral for the Dirac system
$\breve H^D_n$.
 It is the Darboux-dressed momentum operator
of the free  Dirac massive particle (\ref{H0DC}).
The interesting  point is that
for the reflectionless Dirac system 
$\breve H^D_n$ 
 one can identify an additional integral of motion 
$\breve\Gamma$, which satisfies the identity 
$\breve\Gamma^2=1$,
and anticommutes with 
$ \breve{\mathcal{P}}_{n,1}$. 
As a consequence,  
the reflectionless Dirac system
$\breve H^D_n$ can be characterized by the proper exotic 
nonlinear supersymmetry.
Indeed, consider the operator 
$ \breve\Gamma=\mathcal{R}   
\sigma_3$, where
$\mathcal{R}$ 
is the operator of reflection in $x$, $\tau_i$ and $\cC$, 
which satisfies the relations
$\mathcal{R}z=-z\mathcal{R}$, $\mathcal{R}^2=1$,
where 
$z=x,\, \tau_i$, or  $\cC$.
Due to the relations
$[\breve\Gamma,\breve{\mathcal{H}}^D_n]=0$
and $\{\breve\Gamma,\breve{\mathcal{P}}_{n,1}\}=0$,
the $\breve{\mathcal{H}}^D_n$ and   $\breve{\mathcal{P}}_{n,1}$ are 
 identified as bosonic and  fermionic operators, respectively.
They 
generate a nonlinear 
$\mathcal{N}=1$ superalgebra 
\begin{equation}
	[\breve{\mathcal{P}}_{n,1}, \breve{\mathcal{H}}^D_n]=0\,,\qquad
	\{\breve{\mathcal{P}}_{n,1},\breve{\mathcal{P}}_{n,1}\}=2 
	\P_{2(2n+1)}(\breve{\mathcal{H}}^D_n)\,,
\end{equation}
where 
\begin{equation}\label{PnDirac}
    \P_{2(2n+1)}(\breve{\mathcal{H}}^D_n)\equiv \left((\breve{\mathcal{H}}^D_n)^2-
    \mathcal{C}^2\right)
    \prod_{j=1}^n\left((\breve{\mathcal{H}}^D_n)^2-
    (\mathcal{C}^2-\kappa_j^2)\right)^2\,.
\end{equation}
The $2(n+1)$ zeros of the polynomial in $\mathcal{H}^D_n$ operator (\ref{PnDirac}) 
correspond to the energies of the singlet states of the reflectionless Dirac system,
 where  $\breve{\mathcal{E}}^\pm_i=\pm \sqrt{\cC^2-\kappa_i}$, $i=1,\ldots, n$, are the 
 energies of the bound states,
 while $\pm\mathcal{C}$ correspond to the two singlet states
 at the edges of the continuous parts of the 
 spectrum\footnote{Besides a bound  state,
 each double  root $\breve{\mathcal{E}}^\pm_i$, $i=1,\ldots,n$,  
  of the polynomial 
 on left hand side of 
 (\ref{PnDirac}) corresponds also to a non-physical 
 eigenstate of $\breve{\mathcal{H}}^D_n$. }.
 In accordance with  (\ref{est}),
 the left and right moving waves 
 in  (\ref{Phieps})  of the scattering sector,
 which correspond to doubly degenerate energy levels
  $ \breve{\mathcal{E}}^\pm(k^2)$ of 
 $\breve{H}^D_n$   are distinguished by the supercharge
 $\breve{\mathcal{P}}_{n,1}$\,: they are its eigenstates 
 of the opposite sign egenvalues.
 Supplementing the integral  $\breve{\mathcal{P}}_{n,1}$
with a  (nonlocal) integral
 $\breve{\mathcal{P}}_{n,2}=i\breve{\Gamma}\breve{\mathcal{P}}_{n,1}$, 
 the $\mathcal{N}=1$ exotic nonlinear supersymmetry
 of the reflectionless Dirac system  $\breve{H}^D_n$
 can be extended to $\mathcal{N}=2$: 
 $\{\breve{\mathcal{P}}_{n,a},\breve{\mathcal{P}}_{n,b}\}=2 \delta_{ab}
	\P_{2(2n+1)}(\breve{\mathcal{H}}^D_n)$.

 Applying the limit $\cC\rightarrow\kappa_n$, we identify 
 the proper exotic supersymmetric structure of $H^D_n$.
 In this case, the zero energy eigenstate  (\ref{Phi0E0}) of $H^D_n$
 is  also the zero mode of the supercharge ${\mathcal{P}}_{n,1}$. 
 In both Dirac reflectionless systems 
  $\breve{H}^D_n$ and $H^D_n$, the 
  supercharges detect 
  all the non-degenerate eigenvalues of the 
  Hamiltonians by annihilating the 
  corresponding eigenstates, 
  which are the  bound states
  and  the states at the edges of the continuous parts of the spectra. 
  Since the zero energy eigenvalue 
  belongs to the spectrum of $H^D_n$
  but is not present  in the spectrum of 
   $\breve{H}^D_n$, 
  the  proper  exotic supersymmetry of  the Dirac system 
  $\breve{H}^D_n$ 
 is of the broken nature,  while that of ${H}^D_n$ corresponds 
 to the unbroken phase. In correspondence with the second relation
 from (\ref{HSDiam}), the limit $\tau_n\rightarrow\infty$ applied to 
 the Dirac system ${H}^D_n$ with the unbroken  proper exotic 
 supersymmetry
 will produce the system $\breve{H}^{\Diamond D}_{n-1}=
 \breve{\mathcal{S}}^\Diamond_{n-1,1}$,
 see Eq. (\ref{HSDiam}),  characterized by the 
 broken exotic supersymmetry.

\section{Discussion and outlook}

We have considered the two related families
of  the $(1+1)$D Dirac reflectionless systems. 
Each such system corresponds to a fermion in a background of 
a multi-soliton solution (of the kink, or kink-antikink type) 
of the mKdV equation.
In one of these two families, the $n$-soliton  potential
$V^D(x)=\Delta_n(x)$ or $-\Delta_n(x)$,
where  $\Delta_n(x)=\Delta_n(x;\kappa_1,\tau_1,
\ldots,\kappa_n,\tau_n,\cC)$, $\cC^2>\kappa_n^2$, is $(2n+1)$-parametric,
while in the second family the potential $V^D(x)$  is 
$2n$ parametric and corresponds to the function $\cW_n(x)$
or $-\cW_n(x)$, where $\cW_n(x)=\cW_n(x; \kappa_1,\tau_1,
\ldots,\kappa_n,\tau_n)$.
From the viewpoint of the associated extended Schr\"odinger systems,
whose matrix $2\times 2$ Hamiltonians are given by a square 
of the corresponding Dirac Hamiltonian  $H^D=i\sigma_2 
\frac{d}{dx}+\sigma_1V^D(x)$, the Dirac potential $V^D(x)$ 
is a superpotential.  The peculiarity of the considered reflectionless families 
is that in the case of the supersymmetric Schr\"odinger systems, 
in addition to the two first order supercharges $H^D$ and $i\sigma_3H^D$,
they  are characterized by the two  supercharges
to be matrix differentials operators of the order $2n$.
Furthermore,  they possess
 two nontrivial bosonic integrals 
to be differential operators of the order 
$2n+1$, which are constructed from the 
Lax-Novikov integrals of the Schr\"odinger subsystems. One of
these two bosonic integrals  is a central charge 
of the exotic nonlinear superalgebra.
The same higher order central charge  can be identified as 
the supercharge (a fermionic generator) 
of the proper exotic nonlinear supersymmetry
of the reflectionless Dirac system. In the case of 
$V^D(x)=\pm \Delta_n(x)$,
the exotic nonlinear supersymmetries of the Schr\"odinger and Dirac systems
are spontaneously broken, and 
the quantity $(\cC^2-\kappa_n^2)>0$  measures  the
scale of the breaking.
The choices $V^D(x)=\pm\cW_n(x)$ correspond, on the other hand, 
to  the unbroken exotic supersymmetries. 
The interesting point is that there exists a limit procedure, admitting 
the interpretation in the context of a soliton scattering,
which  relates 
the two indicated families of the exotic supersymmetric 
reflectionless systems.
One can define  a kind of a topological charge by
a  relation  
$$
q=\frac{1}{2\vert V^D_0\vert }\int^{\infty}_{-\infty} dx\, \frac{dV^D(x)}{dx}\,, 
$$
where $V^D_0=\lim_{x\rightarrow+\infty} V^D(x)$.
The  case of the broken supersymmetry with 
the kink-antikink type potential $V^D(x)=\pm\Delta_n(x)$ is characterized 
then by $q=0$, while
the cases of the kink, $V^D(x)=-\cW_n(x)$, and anti-kink,  $V^D(x)=\cW_n(x)$,
type potentials of  the unbroken exotic supersymmetries correspond to $q=+1$ and
$q=-1$, respectively.
The quantity  $2\vert V^D_0\vert $ gives   the gap that separates 
the upper and lower continuous bands in the spectrum of the Dirac 
systems, and can be treated as a doubled mass parameter   of a fermion
in an external scalar potential. The mentioned supercharge of the
 Dirac system
annihilates all its non-degenerate energy states, and being  
the Dabroux-dressed momentum operator of 
the  free Dirac particle
(zero-soliton case), distinguishes the left- and right- moving eigenstates
corresponding to the doubly degenerate energy values in the continuum 
bands of the spectrum.
\vskip0.1cm

The described transparent potentials $V^D(x)$ appear in many physical applications 
in the form of stationary solutions for inhomogeneous  
fermion condensates. 
Such self-consistent condensates are described by the equations 
\begin{equation}\label{bdg}
    \left( i \partial \!\!\!/ - V^D\right) \psi_{\alpha} =0,
\qquad
    V^D=-g^2 \sum_{\alpha=1}^N
    \sum_{\rm occ} \bar{\psi}_{\alpha}\psi_{\alpha}\,.
\end{equation}
Here the first equation with  a generalized flavour index 
$\alpha=1,\ldots, N$ represents a system of 
 $(1+1)$D Dirac 
equations,  the $\sum_{\alpha=1}^N$  corresponds to summation in degenerate
states,  and $\sum_{\rm occ} $ corresponds to a sum over 
the completely  filled Dirac sea plus a sum over bound states, which usually are 
partially occupied.   Equations (\ref{bdg}) appear particularly in the superconductivity,  
in the Gross-Neveu model, and in the physics of conducting polymers. 
A famous method of solution of (\ref{bdg})
was realized by Dashen, Hasslacher and Neveu  in 
 \cite{{DaHN}},  where this system  of equations was rewritten 
 in terms of  the scattering data for Schr\"odinger potentials
 $U_\pm=\left(V^D\right)^2\pm \f{d}{dx}V^D-\left(V^D_0\right)^2$,
and as a result it was  shown   that  the reflection 
coefficient for both  potentials 
$U_\pm$ 
has to be equal to zero. For some applications of 
this result, see \cite{SchonTh}--\cite{DunneThies}. 
 Using the ideas of supersymmetry, this picture is equivalent to the search 
 of the first order  operators $D$ and $D^\dagger$,
 which intertwine and  factorize corresponding  
 Schr\"odinger reflectionless Hamiltonians,
$H_+=DD^\dag-E_0$ y $H_-=D^\dag D-E_0$.
As we have shown, 
there are only two situations where such a factorization 
is possible.  
\begin{itemize}
\item When $H_+$ and $H_-$  are completely isospectral,
the  $V^D$ corresponds to the Dirac potentials 
characterized by the topological charge $q=0$, which
are  given by inhomogeneous condensates 
 $\pm\Delta_n$ with asymptotic behaviour  
 $\Delta_n\rightarrow\cC$
for $x\rightarrow\pm\infty$.

\item In other possible case,  
the spectra of $H_+$ and  $H_-$ are different in one 
bound state only, 
and inhomogeneous condensate takes here the form 
 $V^D=-\cW_n$ or $V^D=\cW_n$, where 
  $\cW_n\rightarrow\mp\kappa_n$
for  $x\rightarrow\pm\infty$, and the topological charge 
$q$ takes the  values $+1$ or $-1$.

\end{itemize}

On the other hand, the occupation fraction for each non-degenerate state 
defines the energy of the bound states. Using the method of resolvent, 
J. Feinberg showed in 
 \cite{Feinberg}  that for  all static condensates   the following equality is valid\,:
\begin{equation}
	\nu_i=\frac{2}{\pi}\cot ^{-1}\left(
	\frac{\kappa _i}{\sqrt{(V_0^D)^2-\kappa _i^2}}\right),\quad i=1,\cdots n\,,
\end{equation}
where  $\nu_i$ can take the values  $\nu_i=0,\f{1}{N},\cdots
\f{N-1}{N},1$. This result was reproduced  in  \cite{TakNit} 
for complex kinks in the context of  the 
Bogoliubov-de Gennes and chiral Gross-Neveu systems.

The case $N=1$, $\nu=0,1$ corresponds here to the superconductivity. With these restrictions, 
the topologically trivial  homogeneous condensate is possible,
$V^D=\pm\Delta_0=\pm V^D_0$, $\nu_1=1$ (free Dirac massive particle),    
as well  as the topologically nontrivial inhomogeneous condensate, 
$V^D=\pm\cW_1$, $\nu_1=0$, $\kappa_1=V^D_0$.

The case  $N=2$, $\nu=0,1/2,1$, 
corresponds to polymer conductors in the context of the 
Takayama-Lin-Liu-Maki model \cite{TLM}; in addition to 
$V^D=\pm\Delta_0,\pm\cW_1$,  also  the case $\nu_1=1/2$, 
$V^D=\pm\Delta_1$, $\kappa_1=\frac{1}{\sqrt{2}} \vert V^D_0\vert $
is possible. This last solution is known as a polaron. 
Other topological solution, which is 
kink+polaron 
(or antikink+polaron) corresponds to
 $V^D=\mp\cW_2$ ($\kappa_1=\frac{1}{\sqrt{2}} \vert V^D_0\vert $ and  $\kappa_2=
 \vert V^D_0\vert $).

In the  't Hooft limit 
$N\rightarrow\infty$,   the  $\kappa_i$  can take any value in the interval 
 $0\leq \kappa_i\leq \vert V^D_0\vert $, that makes possible to have any stationary soliton 
 solution. So, we see that for the Gross-Neveu model, the Darboux transformations 
provide a general method to generate real 
inhomogeneous condensates for  (\ref{bdg}).
\vskip0.2cm

 Equations (\ref{Acal1}) and  (\ref{Acal2})
allow us to obtain  a supersymmetric system described by  the  extended first order matrix 
Hamiltonian composed from 
the two Dirac Hamiltonians.
In such a way we can get two different families of the extended systems. The first one 
realizes the unbroken exotic
 supersymmetry and is given by the  
 Hamiltonian of the form 
\be\label{HDext1}
\mathcal{H}^D=\left(%
    \begin{array}{cc}
     {{H}}^D_n & \mathbf{0}\\
    \mathbf{0} & \breve{{H}}^D_{n-1} 
    \end{array}\right)\,.
\ee
The matrix integrals for (\ref{HDext1}) 
given by the first order  differential operators are
 \be\label{SDext1}
\mathcal{S}^D_1=\left(%
    \begin{array}{cc}
    \mathbf{0} & \mathcal{A}_{n} \\
    \mathcal{A}^\dag_{n} & \mathbf{0} \\
    \end{array}%
    \right),\qquad \mathcal{S}^D_2=i\Sigma_3\mathcal{S}^D_1\,,
\ee
where $\Sigma_3$ is a $4\times4$ diagonal matrix  of the form 
$\Sigma_3=diag\,(1_2,-1_2)$
with $1_2$  to be  the unit  $2\times 2$ matrix.
Another  family is given by the Hamiltonian of the form
\be\label{HDext2}
\breve{\mathcal{H}}^D=\left(%
    \begin{array}{cc}
     \breve{{H}}^D_n\\
    \mathbf{0} & \breve{{H}}^D_{n-1}
    \end{array}\right),
\ee
and its  analogous integrals are 
\be\label{SDext2}
\breve{\mathcal{S}}^{D}_1=\left(%
    \begin{array}{cc} 
    \mathbf{0} & \breve{\mathcal{A}_{n}} \\
    \breve{\mathcal{A}}^\dag_{n} & \mathbf{0} \\
    \end{array}%
    \right),\qquad  
    \breve{\mathcal{S}}^{D}_2=i\Sigma_3
    \breve{\mathcal{S}}^{D}_1\,.
\ee 
The grading operator  $\Gamma=\Sigma_3$ identifies the extended Dirac 
 Hamiltonians to be bosonic generators, while
 (\ref{SDext1}) and (\ref{SDext2})  
 are identified as the fermionic generators.
 Then we find that the indicated operators satisfy the  
 nonlinear supersymmetry relations to be of the 
 order $2$ in corresponding Hamiltonians,
$\{\mathcal{S}^D_a,\mathcal{S}^D_b\}=2\delta_{ab}
(\mathcal{H}^{D})^2$,
and 
 $\{\breve{\mathcal{S}}^{D}_a,\breve{\mathcal{S}}^{D}_b\}=
 2\delta_{ab}\left((\breve{\mathcal{H}}^D)^2-\cC^2+\kappa_n^2\right)$.
Besides, in each of the two cases, there exist 
bosonic integrals to be the matrix  differential operators of the order 
$2n+1$, and fermionic integrals of the order $2n$.

Also, it is possible to construct supersymmetric Dirac type systems  with nonlinear 
superalgebraic relations  of the form 
$\{S,S\}=2f((\mathcal{H}^D)^2)$, where $f$ is a polynomial, 
by taking in extended Hamiltonian $\mathcal{H}^D$ a pair of reflectionless Dirac 
Hamiltonians with distinct scattering data. The picture has to be 
 similar to  that obtained in \cite{PAM} 
 for the reflectionless Schr\"odinger systems.

We are going to present the detailed investigation
 of such supersymmetric  pictures with 
extended Dirac Hamiltonians elsewhere.

Note also  that the last relations in (\ref{Acal1}) and (\ref{Acal2}) can be used 
to construct further supersymmetric extensions of the reflectionless
Schr\"odinger systems, in particular, given by $4\times 4$ matrix Hamiltonians.

\vskip0.2cm

\noindent \textbf{Acknowledgements.}
The work has been partially supported by FONDECYT Grant
No. 1130017.   A. A. also acknowledges 
the  CONICYT  scholarship 21120826.


\end{document}